\documentclass[aps,prl,reprint,superscriptaddress,floatfix,raggedbottom]{revtex4-2}

\usepackage[T1]{fontenc}
\usepackage[utf8]{inputenc}
\usepackage{lmodern}
\usepackage{graphicx}
\usepackage{amsmath,amssymb,amsthm,mathtools,bm}
\usepackage{microtype}
\usepackage{xcolor}
\definecolor{red}{RGB}{190,0,0}
\usepackage[colorlinks=true,citecolor=blue!50!black,linkcolor=blue!50!black,urlcolor=blue!50!black]{hyperref}
\usepackage{comment}

\newtheorem{theorem}{Theorem}

\newtheorem{proposition}{Proposition}

\newcommand{\E}{\mathbb E}
\newcommand{\Prob}{\mathbb P}
\newcommand{\R}{\mathbb R}

\newcommand{\dist}{\operatorname{dist}}
\newcommand{\erfc}{\operatorname{erfc}}
\newcommand{\dd}{\mathrm d}
\newcommand{\hs}{\mathrm{hs}}

\begin{document}

\title{Extreme First-Passage Time of Many Interacting Particles}
\author{Ruicheng Bao}
\email{Corresponding author: ruicheng@g.ecc.u-tokyo.ac.jp}
\affiliation{Department of Physics, Graduate School of Science, The University of Tokyo, Hongo, Bunkyo-ku, Tokyo 113-0033, Japan}

\begin{abstract}
Extreme first-passage events are broadly relevant to biological, chemical, and physical processes in which the first successful arrival determines the outcome. Existing theories are largely confined to noninteracting searchers. Interacting extreme-statistics problems are notoriously difficult because correlations destroy probability factorization. We establish a general framework for interacting extreme search. A no-go theorem shows that broad classes of short-range interactions cannot beat the $1/\ln N$ extreme timescale of $N$ independent Brownian searchers, and complementary upper bounds prove that this scale is exact for %broad classes of 
repulsive interactions. We then identify three sharp mechanisms beyond the logarithmic class and derive a unified interaction-driven acceleration limit. In particular, deterministic pairwise interaction can at most reduce the extreme search time to order $1/N$, while stochastic pairwise forcing attains $1/(N\ln N)$. Our results separate acceleration due to statistical redundancy from that generated by coherent many-body transport or amplified fluctuations, deepening our understanding of interacting %stochastic
many-body systems.
\end{abstract}

\maketitle

\textit{Introduction---}First-passage observables quantify when a noisy system reaches a prescribed set for the first time. They provide fundamental timescales across physics, chemistry, biology, ecology, and information processing, governing when transport, reaction, switching, or search is completed. Representative examples include cellular activation, diffusion-controlled reactions, transport in confinement, and search in complex media \cite{Redner2001,benichou2007first,benichou2008enhanced,BenichouVoituriez2014,benichou2016mean,bao2024nonlinear,bao2026exact}. 

In many cases, the \textit{extreme} first-passage time (EFPT) is more physically and biologically relevant. It asks for the earliest arrival among many searchers. EFPT is the appropriate measure of efficiency whenever one successful searcher triggers the response, as in %redundancy-based 
signaling
\cite{Schuss2019,BasnayakeHolcman2020Cusp,
PaquinLefebvre2022Redundancy},
ligand--receptor activation
\cite{HolcmanSchuss2005Microdomains,
DOrsognaChou2005Ligand,DaoDucHolcman2010Threshold}, calcium transduction in cells \cite{Basnayake2019}, decision making \cite{24pre_decision}, narrow escape \cite{18prl_narrow} and fertilization \cite{Ikawa2010,redner15_prl,LawleyUniversal2020}. 

One line of research on EFPT examines the effects of interactions on
escape problems \cite{18prl_narrow,KumarPalShpielberg2024,KumarPalShpielberg2025, BiswasMajumdarPal2025}. Another important direction concerns the role of particle number: increasing the number of searchers can dramatically reduce the EFPT, a mechanism often called the redundancy principle \cite{redner15_prl,Schuss2019} in biological systems. The earliest theoretical analysis of the redundancy-induced acceleration dates back roughly four decades, when a pioneering study \cite{weissOrderStatisticsFirst1983} derived the celebrated $1/\ln N$ scaling for the EFPT of $N$ identical and independent one-dimensional Brownian searchers in the large-$N$ limit. In recent years, the subject has witnessed a renewed surge of interest, prompting a series of deeper investigations into the EFPT scaling of many searchers \cite{redner15_prl,Schuss2019,BasnayakeHolcman2020Cusp,Grebenkov2020,LawleyUniversal2020,LawleyDistribution2020,LawleyMadridEscape2020,MadridLawley2020,linn22jpa,lawleyCompetitionManySearchers2023,24pre_dynamic,25prr_injection,lawley_2025immigration,SposiniPRL,SposiniPRE,24pre_corwin,24prl_corwin,linn26dynamic,26chaos_bigjump,26pre_lawley,26arxiv_bounded,26arxiv_finite_speed}. In particular, the theory has since been extended to the full EFPT distribution \cite{LawleyDistribution2020}, arbitrary spatial dimensions with drift and space-dependent diffusivity \cite{LawleyUniversal2020}, systems with time-dependent \cite{25prr_injection} or fluctuating numbers of searchers \cite{24pre_dynamic,lawley_2025immigration,linn26dynamic}, multiple targets \cite{linn22jpa}, fluctuating diffusivity \cite{SposiniPRL,SposiniPRE}, crowded cellular
environments \cite{BasnayakeHolcman2020Cusp}, random environments \cite{24prl_corwin,24pre_corwin}, bounded-speed dynamics \cite{26arxiv_bounded,26arxiv_finite_speed,26pre_lawley}, and heavy-tailed noise \cite{26chaos_bigjump}.

However, these theories of redundancy-driven acceleration only concern noninteracting searchers,
leaving the role of interparticle interactions largely unexplored. Realistic systems are rarely noninteracting: steric effects,
crowding and volume exclusion, alignment, Coulomb interactions, and couplings mediated by the surrounding
medium are often essential in living and soft-matter systems
\cite{HoflingFranosch2013Crowding,Marchetti2013ActiveMatter,
Likos2001EffectiveInteractions,Levin2002Electrostatic}. Interactions destroy the factorization of the many-body survival probability into a product of one-searcher survival probabilities, making interacting extreme first passage intrinsically many-body. The central questions are therefore natural: are there any universal laws governing the extreme first-passage statistics of multiple searchers with generic interactions, and is it possible to exploit interactions to beat the fastest-search scale $1/\ln N$ of non-interacting searchers?

Here we answer these questions for a broad class of interacting overdamped systems. First, we prove a no-go theorem showing that {for general short-range interactions, a sufficiently low initial local particle-number density rules out the search faster than the Brownian logarithmic scale}. Second, we identify physical routes to a matching upper bound and thereby obtain an exact logarithmic class. Third, we construct three concrete examples that escape this class: coherent force accumulation, stochastic pairwise kicks, and long-range interaction. Finally, we formulate a unified acceleration limit that incorporates the mechanisms behind these examples. %It yields a $N^{-p}$ speed limit for bounded $(p+1)$-body force channels and the additional logarithmic gain produced by many-body noise. %The {\color{red}first} two explicit models attain the corresponding pairwise cases of these branches.

\textit{Setup---}We consider $N$ interacting overdamped searchers in a $d$-dimensional space. The position $X_i^N(t)\in\R^d$ of the $i$-th searcher at time $t$ follows the overdamped Langevin equation
\begin{equation}
\dd X_i^N(t)=b_i^N(X^N(t))\,\dd t+\sqrt{2D}\,\dd W_i(t),
\qquad i=1,\dots,N,
\label{eq:model}
\end{equation}
with $D>0$ and independent $d$-dimensional Brownian motions $W_i$. The drift term $b_i^N(X^N(t))$ incorporates general interaction among searchers, which may be higher-order. The target set $\Gamma\subset\R^d$ is absorbing, and our central focus is the EFPT defined as
\begin{equation}
T_N:=\inf\{t>0:\exists i\text{ such that }X_i^N(t)\in\Gamma\}.
\label{eq:TN}
\end{equation}
Throughout the paper, we assume an order-one initial target distance,
\begin{equation}
\ell_i^N:=\dist(X_i^N(0),\Gamma),
\qquad
\min_{1\le i\le N}\ell_i^N\ge\ell_0>0,
\label{eq:ell0}
\end{equation}
where $\dist(X_i^N(0),\Gamma)$ denotes the minimum distance from the searcher $i$ to the target, {and $\ell_0$ is independent of $N$}.

\textit{A general no-go theorem---}For $N$ independent, identically distributed searchers started at the common target distance $\ell_0$ with one-particle survival probability $Q_1(t)$, one has the exact factorization $\Prob(T_N>t)=Q_1(t)^N$, and inserting the Brownian short-time tail immediately selects the logarithmic scale $\mathbb ET_N\sim \ell_0^2/(4D\ln N)$  given that the one-particle mean
first-passage time is finite. None of that product algebra survives once the searchers interact.
The natural question is therefore what universal statement remains true with interaction. The theorem below shows that a broad bounded-force class cannot be faster than the Brownian logarithmic benchmark.

\begin{theorem}[No-go theorem]
\label{thm:nogo}
Assume \eqref{eq:model}--\eqref{eq:ell0}, and let $\Gamma\subset\R^d$ be closed and convex. Let $B_{N,t}\ge0$ be deterministic and let $H_{N,t}$ be any event such that on $H_{N,t}$ one has
\[
\sup_{1\le i\le N}\sup_{0\le s\le t\wedge T_N}|b_i^N(X^N(s))|\le B_{N,t}
\quad\text{and}\quad B_{N,t}t<\ell_0.
\]
Then
\begin{equation}
\Prob(T_N\le t)
\le
\Prob(H_{N,t}^c)
+N\,\erfc\!\left(\frac{\ell_0-B_{N,t}t}{\sqrt{4Dt}}\right),
\label{eq:nogo-prob}
\end{equation}
and hence, using Markov's inequality $\E T_N\ge t\Prob(T_N>t)$,
\begin{equation}
\E[T_N]\ge t\Biggl[1-\Prob(H_{N,t}^c)-N\,\erfc\!\left(\frac{\ell_0-B_{N,t}t}{\sqrt{4Dt}}\right)\Biggr]_+.
\label{eq:nogo-mean}
\end{equation}
Here $[x]_+:=\max\{x,0\}$. In particular, if for every fixed $c<\ell_0^2/(4D)$ the displayed force supremum, evaluated at $t=c/\ln N$, is $o(\ln N)$ in probability, then
\begin{equation}
\liminf_{N\to\infty}(\ln N)\,\E[T_N]\ge \frac{\ell_0^2}{4D}.
\label{eq:nogo-main}
\end{equation}
\end{theorem}
The theorem permits the drift budget $B_{N,t}$ to grow with $N$. On the logarithmic window $t_N=c/\ln N$, {$B_{N,t_N}=O(\ln N)$ with $\Prob(H_{N,t_N}^c)\to0$ still yields} a $1/\ln N$ lower bound, whereas the sharp Brownian constant $\ell_0^2/(4D)$ is recovered when $B_{N,t_N}=o(\ln N)$.
The proof, given in the Supplemental Material (SM) \cite{supplemental_material}, {Sec.~S2 A}, is a genuine many-body argument. Its core mechanism is clear. For each particle, fix the unit vector from its initial position toward its nearest point on the convex target. If that particle hits the target, its displacement along this fixed direction must be at least its initial target distance.
On any event where the force stays bounded by $B_{N,t}$ up to time
$t\wedge T_N$, the motion projected onto this targetward direction
is controlled by a one-dimensional Brownian motion with drift budget
$B_{N,t}$, and the reflection principle plus a union bound yield
Eq.~\eqref{eq:nogo-prob}. %No factorization over particles is used.

The physical meaning is transparent and revealing. What matters is the cumulative drift accumulated before the earliest hit. On the logarithmic fastest-search window, an $O(\ln N)$ force budget creates at most an $O(1)$ displacement, while every particle still starts an order-one distance from the target. That is why the Brownian logarithmic timescale survives. Moreover, sublogarithmic force budgets produce a vanishing correction
and thus recover the Brownian constant.

It may not yet be clear which physical situations satisfy
the conditions of Theorem \ref{thm:nogo} to render it applicable. The following proposition shows that, for a broad class of short-range
interactions, these conditions translate into a requirement on the initial
local particle-number density, thus ensuring its wide applicability. Because the theorem only probes the logarithmic short-time window, the relevant input can be verified directly from the initial state: one merely needs to exclude an initially overcrowded interaction neighborhood around any given particle. %{\color{red}The following statement distinguishes finite-range soft forces from the singular Lennard-Jones (LJ) and Weeks-Chandler-Andersen (WCA) forces.}

\begin{proposition}[Short-range interacting systems with low initial density]
\label{cor:degree-main}
Under the conditions of Theorem~\ref{thm:nogo}, consider the pairwise model
\begin{equation}
\dd X_i^N(t)=\sum_{j\ne i}F(X_j^N(t)-X_i^N(t))\,\dd t+\sqrt{2D}\,\dd W_i(t),
\label{eq:pairwise-model}
\end{equation}
\emph{(i) Finite-range soft forces.} Let $F(z)=0$ for $|z|>R$ and $|F(z)|\le K$. If the \textbf{initial configuration} is such that every particle has at most $M_N$ neighbors inside the interaction range $R$, and $M_N=O(\ln N)$, then the logarithmic no-go scale survives,
\begin{equation}
\E[T_N]=\Omega((\ln N)^{-1}).
\label{eq:cor-degree-main}
\end{equation}
If in fact $M_N=o(\ln N)$, then the sharp Brownian constant is recovered, namely, Eq. \eqref{eq:nogo-main} holds.

\emph{(ii) General short-range interactions.} For pair interactions with sufficiently fast-decaying  tails,
under the conditions of SM Theorem~S1 (Sec.~S3), including standard Lennard-Jones (LJ) and Weeks-Chandler-Andersen (WCA) forces with fixed parameters in $d=1,2,3$, suppose the initial positions of any two particles do not completely overlap and each particle has at most $M=O(1)$ neighbors within a fixed radius $R_0>0$. Then Eq.~\eqref{eq:cor-degree-main} holds, even if the minimum pair separation tends to zero with $N$. %This case establishes the scale with a conservative positive coefficient.
\end{proposition}

Case (i) of this proposition follows from Theorem~\ref{thm:nogo} (see SM Sec.~S2 B), and case (ii), where the force itself can be unbounded, from an adapted form of the theorem (End Matter A, SM Sec.~S3). Crucially, both cases require the initial local number density not to be excessively high. Without such an initial crowding restriction, a particle can have $O(N)$ neighbors and acquire a correspondingly large drift, as in Example A.

%{\color{red}For soft forces, the initial count controls the short-time force by a stopping argument (SM Sec.~S2 B). For LJ and WCA, reciprocal force cancellation within finite groups controls displacement (SM Sec.~S3). The LJ result includes its attractive tail; WCA has finite range and is purely repulsive. }

The proposition also covers many nonequilibrium models because of the flexibility of the drift term. Bounded external or self-propulsion terms, conservative or not, simply add to the finite-time drift budget, so active Brownian and run-and-tumble particles with fixed propulsion speed have the same $N$-scaling. Active Brownian Ornstein-Uhlenbeck particles can be handled by the generalized Theorem 1 in End Matter, because their integrated active displacement is $o(1)$ with overwhelming probability on the logarithmic window (see SM Sec.~S2 C).

Notably, the common-$D$ and convex-target assumptions are only for simplicity. A bounded heterogeneous extension with $D$ replaced by a deterministic upper bound $D_{\max}$ and a generalized version of Theorem 1 adapted to any closed target are stated in the End Matter. 

\textit{Upper bounds and the logarithmic class---}Here, we establish a matching logarithmic upper bound when a growing number $m_N\to\infty$ of searchers start within
an $N$-independent distance of a flat target \footnote{We provide a counterexample of the logarithmic upper bound when this ``$N$-independent initial distance'' condition is not satisfied. For independent Brownian searchers on
$z>0$ with target $z=0$ and initial positions $z_i(0)=i\ell$,
$\Prob(T_N>t)\to\prod_{i=1}^{\infty}
{\rm erf}(i\ell/\sqrt{4Dt})>0$. Hence $\E T_N$ approaches a positive
constant of order $\ell^2/D$, rather than vanishing. This constant
can be made arbitrarily large by increasing $\ell$}.
Let the target be the plane $z=0$, let $z>0$ denote the perpendicular
distance from the target, and suppose that $m_N$ candidate labels
satisfy $0<z_i(0)\leq\ell_+$, where $\ell_+$ is independent of $N$.
Two distinct physical settings are sufficient. If the searchers
are confined in a fixed slab (e.g. by a reflecting boundary), the
drift may point in either direction and can arise from arbitrary
interactions. The only requirement is that its outward component grow no faster than logarithmically in $m_N$, with an admissible prefactor.
Without confinement, one instead imposes another
condition: the interactions are reciprocal and purely repulsive,
with no restriction on their strength, while the remaining one-body drift must point toward the target or vanish \footnote{A counterexample exists: on the open half-line, independent particles with constant outward drift have a positive probability never to hit, so $\E T_N=\infty$ for every finite $N$, despite a high-probability hit on the first logarithmic window.}.

\begin{proposition}[Logarithmic upper bound]
\label{prop:flat-upper-main}
Assume that at least $m_N$ labels satisfy $0<z_i(0)\le\ell_+<\infty$. The bound
\begin{equation}
\limsup_{N\to\infty}(\ln m_N)\,\E[T_N]
\le \frac{\ell_+^2}{4D}
\label{eq:flat-upper-main}
\end{equation}
holds in either of the following settings: \emph{(i)} There is confinement. Before absorption, the $z$-coordinates remain in a fixed slab $0<z<L$ (e.g., a reflecting boundary condition at $z=L$), and the total \textbf{outward} drift (harmful) is bounded above by $B_N=o(\ln m_N)$, with no additional assumptions on interaction and drift direction [$B_N\le \alpha\ln m_N$ with $\alpha L<D$ is sufficient
to obtain a logarithmic-scale upper bound, while the stronger
condition $B_N=o(\ln m_N)$ recovers the sharp constant
$\ell_+^2/(4D)$]; or \emph{(ii)} Confinement is optional. In this case, the normal drift should be targetward or zero, and the interaction should be reciprocal and purely repulsive [in contrast, (i) allows attractive and nonreciprocal interactions]. These restrictions allow
arbitrary interaction strength. In setting \emph{(ii)} one moreover has, for every $t>0$,
\begin{equation}
\Prob(T_N>t)\le
\prod_{i=1}^NQ_i^0(t):=\prod_{i=1}^N\operatorname{erf}\!\left(\frac{z_i(0)}{\sqrt{4Dt}}\right),
\label{eq:product-upper-main}
\end{equation}
where $Q_i^0(t)$ is the survival probability of a single drift-free Brownian searcher.
\end{proposition}

More transparently, pure repulsion needs neither interaction-strength restriction nor
spatial confinement; confined soft interactions with attraction also obey
the bound under an initial-density criterion analogous to
Proposition~\ref{cor:degree-main} (SM Sec.~S4 A). The two proofs use complementary ideas, detailed in {SM Secs.~S4 A--B}. The two settings concern different routes to a logarithmic-scale mean-time upper
bound. A logarithmic short-time drift bound $B_N=\alpha \ln m_N$ with a sufficiently small
prefactor $\alpha$ is sufficient to show that $T_N$ is of order
$1/\ln m_N$ with probability tending to one \cite{supplemental_material}. This alone does not control the mean EFPT, because rare, exceptionally
long-lived trajectories may dominate it or even make it infinite.
Confinement suppresses this residual tail and upgrades the
high-probability estimate to the logarithmic upper bound on $\E [T_N]$.
Without confinement, pure repulsion with targetward or zero
one-body drift instead yields Eq. 
\eqref{eq:product-upper-main}, from which the logarithmic mean bound
follows. { Eq.~\eqref{eq:product-upper-main} implies that repulsion cannot delay the first arrival relative to independent searchers.} %In a slab, a shifted-product barrier also permits bounded outward one-body drift while placing no magnitude bound on the reciprocal pair force.

Combining Proposition~\ref{prop:flat-upper-main} with Theorem~\ref{thm:nogo} gives the exact logarithmic universality class whenever both hypotheses apply. In particular, for $N$ searchers initially arranged in a layer parallel to the
planar target, all at the common perpendicular distance $\ell_0$, finite-range bounded reciprocal repulsion with sublogarithmic initial local coordination satisfies $(\ln N)\E T_N\to\ell_0^2/(4D)$. SM Sec.~S6 gives a numerical demonstration of this prediction in a two-dimensional system with many searchers interacting through a repulsive potential.

\textit{How to escape from the logarithmic class---}The no-go theorem relies on an order-one initial gap, sufficiently small deterministic transport on the logarithmic window, and ordinary Brownian-size fluctuations. A faster law must violate at least one of these ingredients.

The first route is only related to the initial configuration and is therefore not interaction-driven. If the nearest initial gap shrinks with $N$, then the EFPT itself naturally shrinks. Faster scales have already been reported for many independent searchers \cite{Grebenkov2020}. With the minimum initial gap fixed, two genuinely interaction-driven resources remain: coherent drift and enhancement of fluctuations. In what follows, we provide three concrete examples.%Figure~\ref{fig:breakers} displays one solvable example of each.

\begin{figure}
\centering
\includegraphics[width=0.97\linewidth]{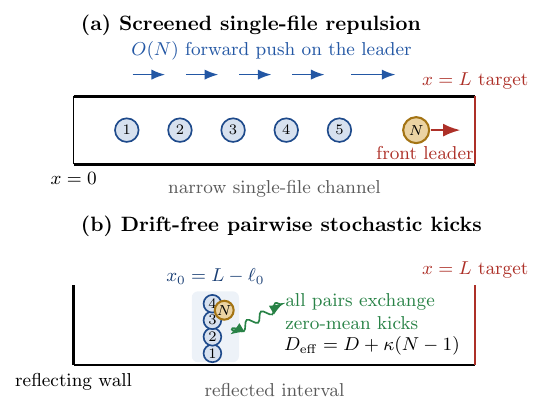}
\caption{Two many-body acceleration mechanisms at fixed initial target distance. (a) In a narrow single-file channel with screened repulsion, the $N-1$ trailing particles generate a coherent $O(N)$ push on the leader, turning the first hit into front propagation. (b) In the drift-free pair-kick model, all pairs exchange equal-and-opposite zero-mean kicks. Each label acquires diffusivity $D+\kappa(N-1)$, and the remaining extreme multiplicity supplies an additional factor $(\ln N)^{-1}$.}
\label{fig:breakers}
\end{figure}

\emph{Example A: screened single-file repulsion.} One-dimensional single-file transport is an effective description of narrow nanopores, ion channels, zeolites, and colloidal microchannels \cite{Wei2000,Karger2021}. In the ordered channel $0\le x_1\le\cdots\le x_N\le L$, with no-passing reflection at
particle contacts [Fig.~\ref{fig:breakers}(a)], take the interaction potential
\begin{equation}
U_N(x)=g\sum_{1\le i<j\le N}e^{-(x_j-x_i)/\lambda},
\qquad g,\lambda>0.
\label{eq:screened}
\end{equation}
If absorption occurs when the leader reaches $L$, then $-\partial_{x_N}U_N\ge(g/\lambda)e^{-L/\lambda}(N-1)$, and the argument detailed in \cite{supplemental_material} yields
\begin{equation}
\E[T_N]\le \frac{\lambda e^{L/\lambda}}{g}\,
\frac{L-X_N(0)}{N-1}=O(N^{-1}).
\label{eq:screened-bound}
\end{equation}
The interaction forces add coherently on one target-facing coordinate.

This example shows that the initial-state control in Theorem~\ref{thm:nogo} is
necessary for preserving the logarithmic scale, rather than a
technical artifact. Notably, the same mechanism can also be realized for the WCA interaction by placing a logarithmic number of
particles within one interaction range of the leader, thereby
generating a superlogarithmic targetward force \footnote{For WCA repulsion, a group of
$K_N=\lceil\ln N\rceil$ particles in the channel of Example A,
with fixed length $L$ below the interaction cutoff, already
has a mean EFPT smaller than
$O(K_N^{-12})=O((\ln N)^{-12})$.}.

\emph{Example B: drift-free pair kicks.} On the interval $[0,L]$ with reflection at $0$ and  absorption at $L$ [Fig.~\ref{fig:breakers}(b)], let
\begin{equation}
\dd X_i=
\sqrt{2D}\,\dd W_i
+\sqrt{2\kappa}\!\left(\sum_{j<i}\dd B_{ji}-\sum_{j>i}\dd B_{ij}\right)
+\dd K_i,
\label{eq:kicks}
\end{equation}
where $\{B_{ij}\}_{1\le i<j\le N}$ are independent standard Brownian motions and $K_i$ is reflection at $0$. The pair kicks are antisymmetric and unbiased, yet each label has quadratic variation $2[D+\kappa(N-1)]t$ \cite{supplemental_material}. If all labels start at $L-\ell_0$, then
\begin{equation}
\lim_{N\to\infty}N\ln N\,\E T_N=\frac{\ell_0^2}{4\kappa}.
\label{eq:kick-sharp-main}
\end{equation}
Thus the stochastic coupling contributes an algebraic factor $N^{-1}$ while extreme multiplicity contributes $(\ln N)^{-1}$. {%\color{red}For $L-\ell_0>0$, the reflecting-wall image correction is exponentially subleading; starting at the reflecting wall leaves the same leading constant.} 
See SM Sec.~S5 C for more details.

{\emph{Example C: unscreened Coulomb repulsion.} In three dimensions, let particles with fixed Coulomb coupling $g>0$ initially form a lattice cloud of fixed spacing $a$, occupying one side of a leading particle. Every fixed-radius neighborhood initially contains $O(1)$ particles, yet the cloud radius $R_N\asymp aN^{1/3}$ gives a targetward force of order $gR_N/a^3=\Theta(N^{1/3})$ on the leader. For a fixed gap $\ell_0$ to a planar target, sufficiently small relative to $a$, the force persists on the relevant short-time window and yields $\E T_N=\Theta(N^{-1/3})$ (SM Sec.~S5 E). The Coulomb force violates the short-range condition in
Proposition~\ref{cor:degree-main}, allowing distant forces to accumulate
coherently and beat the $1/\ln N$ scale.}%Distant forces accumulate coherently, placing this example outside the short-range class of Corollary~\ref{cor:degree-main}.}

\textit{Fundamental acceleration limit---}Here, we show that the preceding examples {fall under} one unified acceleration limit. Fix a test time $t$. Suppose that, except on an event of probability $\varepsilon_N(t)$, the targetward drift accumulated by every searcher before the first hit is at most $B_Nt$, while the quadratic variation of its targetward martingale is at most $2D_N^{\rm eff}s$ up to every $s\le t\wedge T_N$. Here $B_N$ and $D_N^{\rm eff}$ are deterministic short-time budgets that may depend on $N$. A precise formulation is given in End Matter C and SM Sec.~{S5 A}. The acceleration limit is given by the following theorem.

\begin{theorem}[Fundamental acceleration limit]
\label{thm:acceleration}
For a
closed convex target, if the preceding short-time conditions on the targetward drift and the quadratic variation hold at $t=t_N^\star$ with $B_Nt_N^\star<\ell_0$, then
\begin{equation}
\begin{aligned}
&\E T_N\ge t_N^\star[1-o(1)],\\
& t_N^\star\equiv \frac{\ell_0^2}{\left(\sqrt{D_N^{\rm eff}\ln N+B_N\ell_0}+\sqrt{D_N^{\rm eff}\ln N}\right)^2},
\end{aligned}
\label{eq:unified-master-main}
\end{equation}
provided $\varepsilon_N(t_N^\star)\to0$.
\end{theorem}

\begin{table}[!htb]
\caption{Sharp branches of Theorem~\ref{thm:acceleration} at fixed initial gap. $p,q>0$.}
\label{tab:branches}
\begin{ruledtabular}
\begin{tabular}{@{}lcc@{}}
Resource & Lower bound & Optimal Example \\
\hline
$B_N=O(N^p)$ drift & $\Omega(N^{-p})$ & A ($p=1$)\\
$D_N^{\rm eff}=O(N^q)$ noise & $\Omega((N^q\ln N)^{-1})$ & B ($q=1$)
\end{tabular}
\end{ruledtabular}
\end{table}

Theorem \ref{thm:acceleration} interpolates continuously between coherent drift and enhanced diffusion, corresponding to the
physical mechanisms in Examples A and B, respectively. {Example C also lies on the deterministic branch: $B_N=O(N^{1/3})$ and $D_N^{\rm eff}=D$ give $\E T_N=\Omega(N^{-1/3})$.} Namely, on the corresponding high-probability short-time windows, $B_N=O(N^p)$ and $D_N^{\rm eff}=O(1)$ imply $\E T_N=\Omega(N^{-p})$, whereas $B_N=O(1)$ and $D_N^{\rm eff}=O(N^q)$ imply $\E T_N=\Omega([N^q\ln N]^{-1})$. For deterministic $(p+1)$-body interactions whose contribution
from each interacting $(p+1)$-tuple is bounded, the force on any given searcher contains at most
$\binom{N-1}{p}=O(N^p)$ interaction terms. Hence
$B_N=O(N^p)$, and the first branch gives the fundamental $N^{-p}$ acceleration limit. The smooth tagged-leader construction in SM Sec.~S5 D exactly attains this scale, so the scale is optimal. Example A  provides the pairwise case $p=1$. For zero-mean $(q+1)$-body stochastic interactions with bounded variance and orthogonal martingale increments, $D_{N}^{\rm eff}=O(N^q)$ and the lower-bound scale is $(N^q\ln N)^{-1}$. Example B attains the pairwise case $q=1$. The two specific regimes corresponding to the two examples are summarized in Table~\ref{tab:branches}. From Theorem \ref{thm:acceleration}, it is also clear that the lower bound is of the order $t^\star_N\asymp 1/\ln N$ when $D_N^{\rm eff}=D$ and $B_N=O(\ln N)$, recovering the logarithmic scale of Theorem \ref{thm:nogo}.

\textit{Conclusion and discussion---}We have developed a general theory of extreme first passage for multiple interacting Brownian searchers. The $1/\ln N$ acceleration for non-interacting systems is a statistical large-number effect: rare Brownian sprints become likely because many candidates are available. Interactions destroy the probability factorization, but Theorem~\ref{thm:nogo} shows that a large class of interactions cannot create a new purely statistical
acceleration of the leading scale and Proposition~\ref{prop:flat-upper-main} shows that general repulsion does not create a leading-scale slowdown either. Together they identify a sharp logarithmic class and specify the physical conditions under which it holds. To obtain a faster scaling one needs a deterministic large-number effect, such as coherent targetward transport supported by the initial geometry, or a change in the diffusive behavior. We provide {three} explicit examples displaying faster scaling, and further derive a universal bound setting the fundamental limit of such interaction-driven acceleration.

The ideas developed here should also extend to interacting many-body systems in broader stochastic settings,
including non-Gaussian, temporally correlated, or heavy-tailed noise \cite{26chaos_bigjump}, bounded-speed dynamics \cite{26arxiv_finite_speed,26arxiv_bounded} and fluctuating diffusivity \cite{SposiniPRL,SposiniPRE}, but those settings require separate short-time tail estimates and will be treated in subsequent work. Nonreciprocal interactions, hydrodynamic couplings, and strong correlations may alter the effective number of particles relevant to the target search. Phase separation, jamming, or collective locking can reduce this effective number to $O(1)$, producing an $N$-independent fastest time or even a time that increases with $N$. Establishing similarly sharp classifications for these slow regimes---as well as other ordered first-passage statistics \cite{barkai25_slowest,LawleyUniversal2020}---is a natural next step. More broadly, the framework and idea may be useful for other interacting extreme-statistics problems. %Another intriguing direction is to investigate the slowest first-passage statistics for interacting systems.

\begin{acknowledgments}
The author is grateful to ChatGPT (OpenAI, versions 5.4–5.6 Pro) for assistance in extending the scope and strengthening the mathematical rigor of a preliminary version of Proposition 2 obtained by the author. ChatGPT (6 Astra Ultra) assisted in developing a rigorous proof of Case (ii) of Proposition 1 based on the author's conjecture. The main framework and all other results were developed by the author. All AI-assisted results were independently verified and refined by the author, who takes full responsibility for the final manuscript. R.B. is supported by JSPS KAKENHI Grant No.
25KJ0766.  
\end{acknowledgments}

\bibliographystyle{apsrev4-2}
\bibliography{refs}

\begin{center}
{\large\textbf{End Matter}}
\end{center}

\medskip
\noindent\textit{A. Arbitrary closed targets and bounded diffusivities.---} The convex projection used in Theorem~\ref{thm:nogo} is not needed. Instead, one can control the full finite-variation displacement. Write
\begin{equation}
\begin{aligned}
X_i^N(t)&=X_i^N(0)+A_i^N(t)
 +\int_0^t\sqrt{2D_i^N(s)}\,\dd W_i(s),\\
0&\le D_i^N(s)\le D_{\max}.
\end{aligned}
\label{eq:end-general-model}
\end{equation}
Here $\Gamma$ is any closed target and $A_i^N$ is continuous, adapted, and of finite variation. For $a\in(0,\ell_0]$ set
\begin{equation}
G_{N,t}(a):=\left\{\max_i\sup_{s\le t\wedge T_N}|A_i^N(s)|\le\ell_0-a\right\}.
\label{eq:end-buffer-good}
\end{equation}
For every $\eta\in(0,1)$ there is $K_{d,\eta}<\infty$ such that
\begin{equation}
\Prob(T_N\le t)\le\Prob(G_{N,t}(a)^c)+NK_{d,\eta}
\exp\!\left[-\frac{(1-\eta)^2a^2}{4D_{\max}t}\right].
\label{eq:end-closed-target}
\end{equation}
Consequently, if $\Prob(G_{N,c/\ln N}(a)^c)\to0$ for every $c<a^2/(4D_{\max})$, then $\liminf_{N\to \infty}(\ln N)\E T_N\ge a^2/(4D_{\max})$. The proof uses a finite net on the sphere and Dambis--Dubins--Schwarz for each projected martingale, see SM Sec.~{S2 D}.

\medskip
\noindent\textit{B. Directional relaxation for convex targets.---} For a closed convex target choose, at each initial point, the targetward supporting direction $e_i^N$ used in Theorem~\ref{thm:nogo}. If
\begin{equation}
G^{\parallel}_{N,t}(a):=\left\{\max_i\sup_{s\le t\wedge T_N}e_i^N\!\cdot A_i^N(s)\le\ell_0-a\right\},
\end{equation}
then, for common diffusivity $D$,
\begin{equation}
\Prob(T_N\le t)\le\Prob((G^{\parallel}_{N,t}(a))^c)+N\erfc\!\left(\frac{a}{\sqrt{4Dt}}\right).
\label{eq:end-directional-master}
\end{equation}
Thus only targetward displacement must be controlled. This statement requires convexity. %or more generally a single supporting halfspace that contains the entire target. 
The proof is in SM Sec.~S2 D.

\medskip
\noindent\textit{C. Proof of Theorem~\ref{thm:acceleration}.---} Let $\Gamma$ be closed and convex, and for each initial point choose a targetward supporting direction $e_i$. Up to $T_N$, write the stopped projected displacement as
\begin{equation}
e_i\!\cdot\![X_i^N(s)-X_i^N(0)]=A_i^N(s)+M_i^N(s),
\label{eq:end-accel-decomp}
\end{equation}
where $A_i^N$ is a continuous finite-variation process and $M_i^N$ is a continuous local martingale with $M_i^N(0)=0$. %This decomposition covers interacting It\^o diffusions with state-dependent or correlated Gaussian noise, as well as reflected diffusions. 
Suppose that on an event $G_{N,t}$,
\begin{equation}
\max_i\sup_{s\le t\wedge T_N}A_i^N(s)\le u_N(t)<\ell_0,
\label{eq:end-accel-budget}
\end{equation}
and define
\begin{equation}
\begin{aligned}
\varepsilon_N(t)&:=\Prob(G_{N,t}^c),\\
\Pi_N(t,a)&:=\max_i\Prob\!\left(G_{N,t}\cap
\left\{\sup_{s\le t\wedge T_N}M_i^N(s)\ge a\right\}\right).
\end{aligned}
\label{eq:end-accel-envelope}
\end{equation}
If $T_N\le t$, convexity ensures that the first particle
to arrive has projected displacement at least $\ell_0$.
On $G_{N,t}$, its noise contribution must therefore reach
$\ell_0-u_N(t)$.
A union bound over the $N$ particles, together with
$\E T_N\ge t\,\Prob(T_N>t)$, gives
\begin{equation}
\begin{aligned}
\Prob(T_N\le t)&\le\varepsilon_N(t)+N\Pi_N(t,\ell_0-u_N(t)),\\
\E T_N&\ge t\,[1-\varepsilon_N(t)-N\Pi_N(t,\ell_0-u_N(t))]_+.
\end{aligned}
\label{eq:end-accel-master}
\end{equation}
If $u_N(t)=B_Nt$ and $\langle M_i^N\rangle_s\le2D_N^{\rm eff}s$ on $G_{N,t}$, Dambis--Dubins--Schwarz and the reflection principle yield $\Pi_N(t,a)\le\erfc(a/\sqrt{4D_N^{\rm eff}t})$. Substitution into \eqref{eq:end-accel-master} yields
\begin{equation}
\begin{aligned}
\Prob(T_N\le t)&\le \varepsilon_N(t)+N\,\erfc\!\left(
\frac{\ell_0-B_Nt}{\sqrt{4D_N^{\rm eff}t}}\right),\\
\E T_N&\ge t\!\left[1-\varepsilon_N(t)-N\,\erfc\!\left(
\frac{\ell_0-B_Nt}{\sqrt{4D_N^{\rm eff}t}}\right)\right]_+.
\end{aligned}
\label{eq:end-unified-bound}
\end{equation}
Theorem~\ref{thm:acceleration} follows by choosing the largest time $t$ for which the bracket in Eq.~\eqref{eq:end-unified-bound} remains asymptotically nontrivial in the $N\to \infty$ limit. Since $\erfc(x)\sim e^{-x^2}/(\sqrt{\pi}x)$, the threshold is determined by
$(\ell_0-B_Nt)^2/(4D_N^{\rm eff}t)=\ln N$,  equivalently
$B_Nt+2\sqrt{D_N^{\rm eff}t\ln N}=\ell_0$.  Its positive solution is exactly $t_N^\star$.  {At $t=t_N^\star$, $N\erfc(\sqrt{\ln N})\to0$, so the bracket tends to one whenever $\varepsilon_N(t_N^\star)\to0$.}

\end{document}

\title{Supplemental Material for ``Extreme First-Passage Time of Many Interacting Particles''}
\author{Ruicheng Bao}
\affiliation{Department of Physics, Graduate School of Science, The University of Tokyo, Hongo, Bunkyo-ku, Tokyo 113-0033, Japan}
\maketitle
\renewcommand{\thefigure}{S\arabic{figure}}
\renewcommand{\thetable}{S\arabic{table}}
\tableofcontents

\section{Setup and notation}
\label{sec:setup}

In this section, we introduce the setup and notation used throughout this Supplemental Material. Recall that our central quantity of interest is the first arrival time among many interacting Brownian searchers.

For each particle number $N$, let $X^N=(X_1^N,\ldots,X_N^N)$ solve
\begin{equation}
\dd X_i^N=b_i^N(X^N)\,\dd t+\sqrt{2D}\,\dd W_i,
\qquad i=1,\ldots,N,
\label{eq:SM-model}
\end{equation}
where $X_i^N\in\R^d$ is the position of particle $i$, $b_i^N$ is its drift velocity, $D>0$ is the thermal diffusivity, and the $W_i$ are independent standard $d$-dimensional Brownian motions. % We assume that Eq.~\eqref{eq:SM-model} has a solution up to the first target hit, i.e., no particle escapes to infinity in finite time. For the singular forces of Secs.~\ref{sec:SM-singular} and~\ref{sec:coulomb}, the existence of such a solution without collisions is justified in those sections.  
For simplicity, initial positions $x_i^N=X_i^N(0)$ are taken to be deterministic. However, our results extend straightforwardly by conditioning to random initial states, independent of the driving noises, provided the assumptions below on the initial configuration hold almost surely. For a nonempty closed absorbing target $\Gamma$, define
\begin{equation}
\tau_i^N=\inf\{t>0:X_i^N(t)\in\Gamma\},\qquad
T_N=\min_i\tau_i^N,\qquad
\min_{1\le i\le N}\dist(x_i^N,\Gamma)\ge\ell_0>0.
\label{eq:SM-TN}
\end{equation}
Here $\dist$ is Euclidean distance, $\ell_0$ is a common lower bound independent of $N$, and $\E T_N$ is the mean extreme first-passage time (EFPT), which may be infinite. We write $\Prob$ and $\E$ for probability and expectation. The survival probability $Q_N(x,t)=\Prob_x(T_N>t)$ evolves under the backward generator
\begin{equation}
\mathcal L_N=\sum_i b_i^N\cdot\nabla_i+D\sum_i\Delta_i,
\end{equation}
where $\nabla_i$ and $\Delta_i$ are the gradient and Laplacian with respect to particle $i$'s position. Here $\Prob_x$ denotes probability for the system starting from the initial configuration $x=(x_1,\ldots,x_N)$. Absorption occurs when any particle reaches $\Gamma$. Every additional absorbing boundary must be included in this stopping event. Explicitly, $\partial_tQ_N=\mathcal L_NQ_N$ for $t>0$, with $Q_N=0$ whenever some $x_i\in\Gamma$ and $Q_N(x,0)=1$ otherwise. %This backward equation motivates the comparison functions of Sec.~\ref{sec:upper}.

Equation~\eqref{eq:SM-model} describes motion in Euclidean space. Reflection contributes an additional finite-variation term and is included explicitly whenever used below. %A periodic coordinate can be treated on its Euclidean lift with the corresponding periodically repeated target. 
Throughout, the diffusivity, interaction parameters, and geometric constants are independent of $N$ unless their dependence is stated. For positive sequences, $f_N=O(g_N)$ means $f_N\le Cg_N$ for all sufficiently large $N$, with $C$ independent of $N$; $f_N=o(g_N)$ means $f_N/g_N\to0$. The notation $f_N=\Omega(g_N)$ means $f_N\ge c g_N$ eventually for an $N$-independent $c>0$; $\Theta(g_N)$ means both this lower bound and a corresponding upper bound, and $f_N\sim g_N$ means $f_N/g_N\to1$. We use $|\cdot|$ for the Euclidean norm of a vector, $s\wedge t=\min(s,t)$, and $E^c$ for the complement of an event $E$. $\mathbf 1_E$ is its indicator, equal to one on $E$ and zero otherwise. The superscript $N$ on particle positions is omitted where the particle number is clear.

\emph{Organization.} Sections~\ref{sec:lower} and~\ref{sec:SM-singular} prove lower bounds on $\E T_N$, showing when the search cannot be faster than the Brownian $1/\ln N$ law. Section~\ref{sec:upper} proves matching upper bounds. Section~\ref{sec:acceleration} quantifies how much faster interactions can make the search and analyzes the examples of the main text, and Sec.~\ref{sec:numerics} reports numerical checks. %Two elementary ideas recur throughout. (i)~\emph{Lower bounds via the first hitter.} The first particle to arrive must have crossed the initial gap. If the directed motion generated by the interactions cannot cover the gap within a short time window, the first hitter's own thermal noise must supply the missing displacement, an event with a Gaussian tail. Summing this tail over the $N$ possible first hitters (a union bound) bounds $\Prob(T_N\le t)$ without ever multiplying probabilities of interacting particles. (ii)~\emph{Upper bounds via independent noises or comparison functions.} Either the independent thermal noises of many candidate particles provide many independent chances to arrive, or a product of single-particle survival probabilities is shown to bound the interacting survival probability from above. A bound on the \emph{mean} EFPT additionally requires control of the rare realizations in which no particle arrives early.

\section{Logarithmic lower bounds (no-go theorem) and the case (i) in Proposition 1}
\label{sec:lower}

In this section, we provide complementary derivations of Theorem 1 and case (i) of Proposition 1 in the main text. We derive conditions under which interactions cannot make the first arrival faster than the Brownian logarithmic scale. The key is that thermal motion must still cover a finite part of the initial target gap. The logarithm is an extreme-value effect: a free searcher crosses a gap $\ell_0$ within a short time $t$ with probability $\simeq e^{-\ell_0^2/(4Dt)}$, so among $N$ searchers such a crossing becomes likely only when $Ne^{-\ell_0^2/(4Dt)}\sim1$, i.e., at $t\simeq\ell_0^2/(4D\ln N)$. Interactions can change this only if, within this short window, they move some particle by a finite fraction of the gap in the large $N$ limit.

\subsection{Proof of Theorem 1 of the main text}
\label{sec:projection}

For a convex target, every arriving particle must move a minimum distance in a suitable direction. Bounding the motion caused by interactions leaves a distance that thermal fluctuations must cover. %Projecting onto this direction turns the $d$-dimensional many-body problem into $N$ one-dimensional problems, which are then treated by a union bound.

Let $\Gamma$ be closed and convex. For each $x_i^N$, choose its nearest target point $z_i$ and the targetward unit vector $e_i=(z_i-x_i^N)/|z_i-x_i^N|$. Convexity gives
\begin{equation}
e_i\cdot(y-x_i^N)\ge\dist(x_i^N,\Gamma)\ge\ell_0,
\qquad y\in\Gamma.
\label{eq:supporting-hyperplane}
\end{equation}
Geometrically, only the plane through $z_i$ perpendicular to $e_i$ is relevant: every target point lies at least a distance $\dist(x_i^N,\Gamma)$ farther along $e_i$ than $x_i^N$ does. To verify Eq.~\eqref{eq:supporting-hyperplane}, take any $y\in\Gamma$. By convexity, the segment $z_i+\theta(y-z_i)$, $0\le\theta\le1$, lies in $\Gamma$. Since $z_i$ is the target point nearest to $x_i^N$, the squared distance $f(\theta)=|x_i^N-z_i-\theta(y-z_i)|^2$ is minimal at $\theta=0$, so $f'(0)=-2(x_i^N-z_i)\cdot(y-z_i)\ge0$, i.e., $e_i\cdot(y-z_i)\ge0$. Adding $e_i\cdot(z_i-x_i^N)=|z_i-x_i^N|=\dist(x_i^N,\Gamma)$ gives Eq.~\eqref{eq:supporting-hyperplane}. Consequently, particle $i$ can reach $\Gamma$ only if its displacement along the fixed direction $e_i$ is at least $\ell_0$. This necessary condition is one dimensional: instead of the full $d$-dimensional path, it suffices to follow one scalar coordinate per particle, $e_i\cdot X_i^N(t)$.

Suppose an event $H_{N,t}$ satisfies $\max_i\sup_{s\le t\wedge T_N}|b_i^N(X^N(s))|\le B_{N,t}$, where $B_{N,t}\ge0$ is a deterministic drift bound and $B_{N,t}t<\ell_0$. We call $H_{N,t}$ a good event: on it, no particle feels a drift larger than the budget $B_{N,t}$ up to time $t\wedge T_N$. Later times do not matter, because the dynamics stops at the first hit. If particle $i$ makes the first hit by time $t$, then on $H_{N,t}$,
\begin{equation}
\sqrt{2D}\,e_i\cdot W_i(T_N)\ge\ell_0-B_{N,t}t.
\label{eq:SM-first-hitter}
\end{equation}
To see this, write $X_i^N(T_N)-x_i^N=\int_0^{T_N}b_i^N\,\dd s+\sqrt{2D}\,W_i(T_N)$ and project onto $e_i$: the left side is at least $\ell_0$ by Eq.~\eqref{eq:supporting-hyperplane}, and the drift term is at most $B_{N,t}t$ on $H_{N,t}$. In words, the interactions can shorten the distance that the noise must cover from $\ell_0$ to $\ell_0-B_{N,t}t$, but not to zero. Because $e_i$ is fixed at time zero, $\beta_i(s)=e_i\cdot W_i(s)$ is a standard one-dimensional Brownian motion. Equation~\eqref{eq:SM-first-hitter} thus requires the one-dimensional Brownian motion $\sqrt{2D}\beta_i$ to reach the level $u=\ell_0-B_{N,t}t$ by time $t$, and the reflection principle \cite{KaratzasShreve} gives the exact probability of this event, $\Prob[\sup_{s\le t}\sqrt{2D}\beta_i(s)\ge u]=\erfc(u/\sqrt{4Dt})$. If $T_N\le t$ on $H_{N,t}$, at least one of these $N$ events must occur.
%Choose the smallest label to break ties between simultaneous first hits.  
Choose the particle with the smallest label if they hit at the same time. A \textit{union bound} over these first-hitter events and the one-dimensional reflection principle give
\begin{align}
\Prob(T_N\le t)&\le\Prob(H_{N,t}^c)+N\erfc\!\left(\frac{\ell_0-B_{N,t}t}{\sqrt{4Dt}}\right),
\label{eq:SM-good-event-main}\\
\E T_N&\ge t\left[1-\Prob(H_{N,t}^c)-N\erfc\!\left(\frac{\ell_0-B_{N,t}t}{\sqrt{4Dt}}\right)\right]_+.
\label{eq:SM-good-event-mean}
\end{align}
Here $\erfc$ is the complementary error function and $[u]_+=\max(u,0)$. The second inequality follows from $T_N\ge t\mathbf1_{\{T_N>t\}}$ (Markov's inequality). This proves Eqs. (4)-(5) in Theorem~1 of the main text. %The union bound only adds probabilities, so it holds however strongly the interactions correlate different particles.

For $t=c/\ln N$, suppose that the force supremum $S_N=\max_i\sup_{s\le t\wedge T_N}|b_i^N|$ is $o(\ln N)$ in probability. Then one can choose a deterministic $\varepsilon_N\to0$, decaying slowly enough, such that $H_{N,t}=\{S_N\le\varepsilon_N\ln N\}$ is a good event with budget $B_{N,t}=\varepsilon_N\ln N$ and $\Prob(H_{N,t}^c)\to0$.  Over the window $t=c/\ln N$ the drift then displaces a particle by at most $B_{N,t}t=\varepsilon_Nc\to0$: the interactions have no time to carry any particle across a finite gap. Since $\erfc(y)\sim e^{-y^2}/(\sqrt\pi y)$, Eq.~\eqref{eq:SM-good-event-main} tends to zero for every $c<\ell_0^2/(4D)$. Explicitly, $\erfc(y)\le e^{-y^2}$ gives $N\erfc[(\ell_0-\varepsilon_Nc)/\sqrt{4Dc/\ln N}]\le N^{1-(\ell_0-\varepsilon_Nc)^2/(4Dc)}$, which vanishes when $c<\ell_0^2/(4D)$. This competition between the number $N$ of candidates and the Gaussian tail $\simeq N^{-\ell_0^2/(4Dc)}$ of each candidate is the origin of the $1/\ln N$ scale. Equation~\eqref{eq:SM-good-event-mean} then gives $\E T_N\ge(c/\ln N)[1-o(1)]$ for every such $c$. Thus
\begin{equation}
\liminf_{N\to\infty}(\ln N)\E T_N\ge\frac{\ell_0^2}{4D}.
\label{eq:SM-sharp-lower}
\end{equation}

More generally, if the good-event budget satisfies $B_{N,c/\ln N}\le(\alpha+o(1))\ln N$ for a fixed $\alpha\ge0$, then the same argument works whenever $\alpha c+2\sqrt{Dc}<\ell_0$. Nothing else in the argument changes, and no further assumption on the dynamics is needed. The only difference is that the drift may now use up the finite distance $\alpha c$ of the gap during the window, so the noise has to supply $\ell_0-\alpha c$ instead of $\ell_0$, and the erfc term in Eq.~\eqref{eq:SM-good-event-main} is at most $N^{1-(\ell_0-\alpha c)^2/(4Dc)+o(1)}$. The condition $\alpha c+2\sqrt{Dc}<\ell_0$ has a direct reading: the drift displacement $\alpha c$ plus the largest Brownian excursion among $N$ searchers, $\sqrt{4Dt\ln N}=2\sqrt{Dc}$, must remain smaller than the gap. Solving $\alpha c+2\sqrt{Dc}=\ell_0$ for its positive root gives the limiting lower constant
\begin{equation}
c_-(\alpha)=\frac{\ell_0^2}{(\sqrt{D+\alpha\ell_0}+\sqrt D)^2}>0,
\label{eq:SM-calpha}
\end{equation}
that is, $\liminf_{N\to\infty}(\ln N)\E T_N\ge c_-(\alpha)$, provided $\Prob(H_{N,c/\ln N}^c)\to0$ for each $c<c_-(\alpha)$. Thus an $O(\ln N)$ force budget only replaces the Brownian prefactor $c_-(0)=\ell_0^2/(4D)$ by $c_-(\alpha)$. This prefactor decreases with $\alpha$ but remains strictly positive for every finite $\alpha$. An upper bound of order $\ln N$ on the forces is therefore already sufficient for the $1/\ln N$ lower bound. For $\alpha\ell_0\gg D$ one has $c_-(\alpha)\simeq\ell_0/\alpha$, i.e., $t\simeq\ell_0/B_{N,t}$ is simply the time the drift alone would need to cross the gap. Equation~\eqref{eq:SM-calpha} is a special case of the general acceleration limit of Sec.~\ref{sec:accel-master}.

\subsection{Case (i) of Proposition 1: Initial-density constraint for soft forces}
\label{sec:soft}

For a system with bounded finite-range force, the total drift is controlled by \emph{how many neighbors contribute}. The initial neighbor count bounds the force until a prescribed displacement occurs, and that displacement itself requires a rare thermal excursion. This gives a criterion solely based on the initial preparation, as shown below. The point is that Theorem~1 needs a force bound along the trajectory up to the first hit, whereas Proposition~1 assumes something only about the initial configuration. The link is a stopping argument, which allows Theorem~1 to be applied directly: as long as no particle has moved farther than a fixed distance $\delta$, every neighborhood, and hence every force, is controlled by the initial one. And moving by $\delta$ early is itself a rare thermal event (it plays the role of the ``bad event'' $H_{N,t}^c$).

Retain the convex target of Sec.~\ref{sec:projection}. Consider $b_i^N=\sum_{j\ne i}F(X_j^N-X_i^N)$ with $|F(z)|\le\Phi(|z|)$, where $\Phi$ is a nonnegative, nonincreasing bound on the pair-force magnitude. For a displacement radius $\delta>0$, define the initial force envelope
\begin{equation}
\mathcal B_N(\delta)=\max_i\sum_{j\ne i}\Phi\bigl((|x_i^N-x_j^N|-2\delta)_+\bigr).
\label{eq:SM-BNdelta}
\end{equation}
Until a particle moves by $\delta$, all pair distances are at least their initial values minus $2\delta$ (triangle inequality), so every drift is bounded by $\mathcal B_N(\delta)$, because $\Phi$ is nonincreasing. Let $\sigma_\delta$ be the first time at which some particle has moved by $\delta$ from its initial position. Then $H_{N,t}=\{\sigma_\delta>t\wedge T_N\}$ is a good event in the sense of Theorem~1, with the deterministic budget $B_{N,t}=\mathcal B_N(\delta)$, so Eq.~\eqref{eq:SM-good-event-main} applies without modification. Now, it only remains to bound $\Prob(H_{N,t}^c)$. If this displacement threshold is reached before $t\wedge T_N$, some Brownian driver has norm at least $(\delta-\mathcal B_N(\delta)t)/\sqrt{2D}$.

To bound the probability of this $d$-dimensional excursion, we need a Gaussian tail for the norm of a Brownian motion, not only for a fixed projection as in Sec.~\ref{sec:projection}. For every $\eta\in(0,1)$ there is a finite $N$-independent constant $K_{d,\eta}$ such that
\begin{equation}
\Prob\left(\sup_{s\le t}|\sqrt{2D}W(s)|\ge r\right)
\le K_{d,\eta}\exp\left[-\frac{(1-\eta)^2r^2}{4Dt}\right].
\label{eq:SM-radial-sup}
\end{equation}
To see this, cover the unit sphere by finitely many directions $u$ so that some projection obeys $u\cdot v\ge(1-\eta)|v|$, then apply the reflection principle to each projection. The prefactor $K_{d,\eta}$ and the small loss $\eta$ in the exponent are the price for controlling all directions at once. Neither affects the leading asymptotics, because $\eta$ can be taken arbitrarily small. Consequently, if $\mathcal B_N(\delta)t<\min(\delta,\ell_0)$,
\begin{align}
\Prob(T_N\le t)\le{}&NK_{d,\eta}
\exp\left[-\frac{(1-\eta)^2(\delta-\mathcal B_N(\delta)t)^2}{4Dt}\right]
+N\erfc\left(\frac{\ell_0-\mathcal B_N(\delta)t}{\sqrt{4Dt}}\right).
\label{eq:SM-shell-bound}
\end{align}
Equation~\eqref{eq:SM-shell-bound} is thus Eq.~\eqref{eq:SM-good-event-main} of Theorem~1 with this good event: the second term is its Brownian term with budget $\mathcal B_N(\delta)$, and the first term bounds $\Prob(H_{N,t}^c)$ through Eq.~\eqref{eq:SM-radial-sup}. This stopping argument avoids any assumption about the density at later times. For $\delta>\ell_0$ and $\limsup_N\mathcal B_N(\delta)/\ln N\le\alpha$, both terms vanish at every $c<c_-(\alpha)$ after choosing $\eta$ sufficiently small. Choosing $\delta>\ell_0$ makes an early stopping require a larger excursion than reaching the target itself, so, once $\eta$ is small, the first term decays faster than the second on the logarithmic window. In particular, $\mathcal B_N(\delta)=o(\ln N)$ gives Eq.~\eqref{eq:SM-sharp-lower}.

For $|F|\le K$ and $F=0$ beyond range $R$, suppose each initial particle has at most $M_N$ other particles within distance $R$. Cover the ball of radius $R+2\delta$ by a fixed number $C_{d,R,\delta}$ of sets of diameter at most $R$. Each occupied set contains at most $M_N+1$ particles, because any one of them sees all the others within $R$. Therefore
\begin{equation}
\mathcal B_N(\delta)\le K C_{d,R,\delta}(M_N+1).
\end{equation}
Taking any fixed ($N$-independent) $\delta>\ell_0$ proves case (i) of Proposition~1: $M_N=O(\ln N)$ preserves the logarithmic lower scale, and $M_N=o(\ln N)$ preserves the sharp constant. Reciprocity and a repulsive sign are unnecessary here. In particular, attractive and nonreciprocal soft forces are allowed.

\emph{Bounded forces with infinite range but integrable tails.} Notably, the finite-range cutoff is not necessary. Suppose each initial shell $kR\le|x_j^N-x_i^N|<(k+1)R\ (k=0,1,\ldots)$ contains at most $C\rho_N(1+k)^{d-1}$ particles, with fixed $R,C>0$. Then $\mathcal B_N(\delta)=O(\rho_N)$ whenever $\sum_k(1+k)^{d-1}\Phi((kR-2\delta)_+)$ converges, i.e., whenever $\int_1^\infty r^{d-1}\Phi(r)\,\dd r<\infty$, so that the force decays faster than $r^{-d}$. With $\rho_N$ playing the role of $M_N$, $\rho_N=O(\ln N)$ preserves the logarithmic scale and $\rho_N=o(\ln N)$ the sharp constant. This is exactly the short-range condition~\eqref{eq:SM-singular-tail}, which Sec.~\ref{sec:SM-singular} extends to forces that diverge at contact. For slower tails, $\Phi\propto r^{-\gamma}$ with $\gamma<d$, distant particles can push coherently and beat the logarithmic scale even at bounded density (Sec.~\ref{sec:coulomb}). The case $\gamma=d$ is marginal: at bounded density it only reduces the prefactor to $c_-(\alpha)$. %Bounded extra drift adds its magnitude to $\mathcal B_N(\delta)$.

\subsection{Active particles}
\label{sec:active}

We consider particles that self-propel as well as diffuse thermally. Fixed-speed propulsion and Gaussian active velocities can quantitatively change arrival times while preserving the logarithmic lower bound, even with the Brownian constant $\ell_0^2/(4D)$. The reason is a separation of scales: within the short window $t\sim1/\ln N$, persistent motion displaces a particle only by an amount that vanishes as $N\to\infty$, whereas the rare thermal excursion that produces the first arrival must cover the whole gap.

Active Brownian particles and run-and-tumble particles with fixed speed $v_0$ add a displacement of norm at most $v_0t$. They therefore replace a force budget $B_{N,t}$ by $B_{N,t}+v_0$ in the preceding estimates, irrespective of the orientation dynamics. Since $v_0=o(\ln N)$, the sharp constant of Eq.~\eqref{eq:SM-sharp-lower} is unchanged. A finite speed only affects finite-$N$ corrections, as in the active system of Sec.~\ref{sec:numerics}.

For an active Ornstein--Uhlenbeck velocity, retain the thermal Brownian term with $D>0$ and write
\begin{equation}
\dd U_i=-\tau^{-1}U_i\,\dd t+\sqrt{2\sigma^2/\tau}\,\dd B_i,
\qquad A_i^{\rm act}(s)=\int_0^s U_i(r)\,\dd r,
\end{equation}
where $U_i$ is the active velocity, $A_i^{\rm act}$ is its accumulated displacement, and $B_i$ is a standard $d$-dimensional Brownian motion. The fixed parameters $\tau>0$ and $\sigma\ge0$ set the velocity persistence time and stationary velocity scale. The centered Gaussian initial velocities have uniformly bounded covariance. Unlike a fixed propulsion speed, this active velocity is unbounded, so we cannot simply add it to the force budget. Instead, we show that the accumulated active displacement is uniformly small on the logarithmic window and include it in the finite-variation part $A_i^N$ of Sec.~\ref{sec:buffer}. The explicit solution and integration by parts imply
\begin{equation}
\sup_{s\le t}|U_i(s)|\le |U_i(0)|+C_{\tau,\sigma}\sup_{s\le t}|B_i(s)|.
\end{equation}
Uniform Gaussian tails and Eq.~\eqref{eq:SM-radial-sup} then give, for every fixed $a,c>0$ and $t_N=c/\ln N$,
\begin{equation}
\Prob\left(\max_i\sup_{s\le t_N}|A_i^{\rm act}(s)|>a\right)
\le C_1N e^{-C_2(\ln N)^2}+C_3N e^{-C_4(\ln N)^3}\longrightarrow0.
\end{equation}
All $C_j>0$ are independent of $N$. Thus the active displacement is negligible uniformly over labels on the logarithmic window. With $A_i^{\rm act}$ included in $A_i^N$, Eq.~\eqref{eq:SM-directional}, or Eq.~\eqref{eq:SM-buffered-master} for a general target, applies with any $a<\ell_0$. If the interaction forces are $o(\ln N)$, this yields the sharp constant $\ell_0^2/(4D)$. This conclusion concerns Brownian particles with additional active propulsion, and it uses the nonzero thermal diffusivity $D$. Without thermal noise, the first arrival would instead be controlled by rare large active velocities, which is a different problem not considered here.

\subsection{Generalized Theorem 1 in End Matter: arbitrary closed targets}
\label{sec:buffer}

Up to now, we only consider the convex targets for brevity. Here, we generalize the no-go result to arbitrary closed targets. %Reaching a target requires crossing the initial gap, even when the target is irregular or disconnected. This gives lower bounds for general target shapes and continuous noises that may be correlated between particles. 
For such targets no single direction works for all target points, so we control the full displacement instead of one projection, which costs only a constant prefactor and an arbitrarily small loss in the exponent.

Let $\Gamma$ be any closed target, and suppose
\begin{equation}
X_i^N(s)=x_i^N+A_i^N(s)+M_i^N(s),\qquad A_i^N(0)=M_i^N(0)=0,
\label{eq:SM-semimart}
\end{equation}
where $A_i^N$ is continuous and of finite variation, representing accumulated drift or boundary pushes, and $M_i^N$ is a continuous local martingale representing the fluctuating displacement. This includes integrated drift and explicit reflection in $A_i^N$. It also covers state-dependent, heterogeneous noise. End Matter A is a special case. Assume every unit projection of $M_i^N$ has quadratic variation at most $2D_{\max}s$. Quadratic variation measures the variance accumulated by the martingale, and $D_{\max}>0$ is a deterministic upper bound on the diffusivity. For a fixed particle $i$ and unit direction $e$, set
$m(s)=e\cdot M_i^N(s\wedge T_N)$.
The Dambis-Dubins-Schwarz theorem \cite{RevuzYor} gives
$m(s)=\beta(\langle m\rangle_s)$, where $\beta$ is a
standard one-dimensional Brownian motion.
Since $\langle m\rangle_s\le2D_{\max}s$, the reflection
principle gives, for $b>0$,
\begin{equation}
\Prob\!\left(\sup_{s\le t}m(s)\ge b\right)
\le \erfc\!\left(\frac{b}{\sqrt{4D_{\max}t}}\right)
\le \exp\!\left(-\frac{b^2}{4D_{\max}t}\right).
\end{equation}
Notably, this is equivalent to Eq.~\eqref{eq:SM-radial-sup} with $D=D_{\max}$. Intuitively, a continuous martingale is a Brownian motion running on an internal clock set by its quadratic variation. If this clock never runs faster than $2D_{\max}$, the martingale cannot make larger excursions than a Brownian motion with diffusivity $D_{\max}$.

For $0<a\le\ell_0$, define $G_{N,t}(a)=\{\max_i\sup_{s\le t\wedge T_N}|A_i^N(s)|\le\ell_0-a\}$. A hit on this event requires a martingale displacement of norm at least $a$, since any hit requires $|X_i^N-x_i^N|\ge\ell_0$ while $|A_i^N|\le\ell_0-a$. Fix $0<\eta<1$ and
choose finitely many unit directions $U_{d,\eta}$ such that
$\max_{e\in U_{d,\eta}}e\cdot v\ge(1-\eta)|v|$
for every $v\in\mathbb R^d$.
Such a choice exists by compactness of the unit sphere.
Their number $K_{d,\eta}=|U_{d,\eta}|$ depends only on
$d$ and $\eta$.
Thus a displacement of norm at least $a$ requires a
projection of at least $(1-\eta)a$ in one of these directions.
Applying the preceding bound with $b=(1-\eta)a$ and
summing over the directions and particles gives
\begin{equation}
\Prob(T_N\le t)\le\Prob(G_{N,t}(a)^c)+NK_{d,\eta}
\exp\left[-\frac{(1-\eta)^2a^2}{4D_{\max}t}\right].
\label{eq:SM-buffered-master}
\end{equation}
If the failure probability tends to zero at $t=c/\ln N$ for every $c<a^2/(4D_{\max})$, then $\liminf(\ln N)\E T_N\ge a^2/(4D_{\max})$. This proves End Matter A. Independence between martingales is unnecessary, because only a union bound is used.

For a convex target and common Brownian diffusivity $D$, the weaker event
$G^\parallel_{N,t}(a)=\{\max_i\sup_{s\le t\wedge T_N}e_i\cdot A_i^N(s)\le\ell_0-a\}$ suffices. It restricts only the targetward component of the directed displacement, so arbitrarily strong pushes away from the target or parallel to it are allowed. Using the supporting directions of Eq.~\eqref{eq:supporting-hyperplane} gives
\begin{equation}
\Prob(T_N\le t)\le\Prob((G^\parallel_{N,t}(a))^c)+N\erfc\left(\frac a{\sqrt{4Dt}}\right),
\label{eq:SM-directional}
\end{equation}
which proves End Matter B. Only displacement toward the target enters this estimate. Notably, this argument may not be applied to a  nonconvex target. Eq.~\eqref{eq:SM-buffered-master} should be applied. For example, if the target consists of two balls on opposite sides of a particle, the particle can reach the farther ball while moving away from the nearer one.

\section{Case (ii) of Proposition 1: singular forces}
\label{sec:SM-singular}

Many molecular potentials have a divergent short-distance core. In this section, we tackle this situation. We show that general short-range forces (that may have a singular core) still obey a logarithmic lower bound when the initial local particle count is bounded, without imposing a fixed minimum pair separation. The difficulty is that even a single close neighbor can exert an arbitrarily large force, because the interaction diverges at small separation. %The resolution is Newton's third law: such huge forces act inside tight clusters of a few particles, where they cancel in the cluster's mean motion. They can rearrange a cluster locally, but cannot carry any of its members across the finite gap $\ell_0$ within the short logarithmic window.

The force supremum in the main-text Theorem~1 need not be small for a singular potential, so it cannot be applied directly. An isolated pair at separation $\varepsilon_N$ already has Lennard--Jones (LJ) or Weeks--Chandler--Andersen (WCA) force of order $\varepsilon_N^{-13}$, although its initial coordination number is one. The logarithmic lower bound nevertheless follows from a %uniform
initial number-density bound, through the generalized form of Theorem~1 proved in Sec.~\ref{sec:buffer} (End Matter A): Eq.~\eqref{eq:SM-buffered-master} needs a budget only on the drift displacement $A_i^N(s)=\int_0^sb_i^N\,\dd r$, not on the force itself. %Reciprocity cancels internal forces in a group average. To control every particle, the proof combines a partition into small groups with a maximum over subset averages.

\emph{Strategy of the proof.} (1)~\emph{Grouping.} The particles are partitioned into small groups of at most $M$ members, each group being much farther from all other groups than its own size. (2)~\emph{Forces between groups.} As long as no particle has moved much, these forces are bounded by an $N$-independent constant, thanks to the bounded local count and the integrable force tail. (3)~\emph{Forces inside a group.} They may be huge, but they are reciprocal. The envelope $H_a$ of Eq.~\eqref{eq:SM-singular-envelope} below bounds the most advanced member of a group, and its growth involves only bounded forces. (4)~\emph{Conclusion.} Steps (1)--(3) show that, during an $N$-independent time $t_*$ and on an event of small Brownian displacements, every drift displacement $A_i^N$ stays far below $\ell_0$. This is the good event of Eq.~\eqref{eq:SM-buffered-master}, and the first-hitter bound of End Matter A gives Eq.~\eqref{eq:SM-singular-prob}. Compared with Sec.~\ref{sec:soft}, the forces may now diverge at contact. This is why reciprocity and an $N$-independent local count are required.

\begin{theorem}[Initial-density criterion for singular reciprocal forces]
\label{thm:SM-singular-density}
Consider a continuous, nonexplosive, noncolliding solution from distinct initial points, with locally integrable drift, of
\begin{equation}
\dd X_i=\left[g_i(X)+\sum_{j\ne i}F(X_i-X_j)\right]\dd t
+\sqrt{2D}\,\dd W_i,
\qquad |g_i|\le B_0,
\label{eq:SM-singular-model}
\end{equation}
Here $X=(X_1,\ldots,X_N)$ is the full configuration, $g_i(X)$ is an additional $N$-independent single-particle drift and $F(-z)=-F(z)$ expresses equal and opposite pair forces. The force here takes $X_i-X_j$ as its argument. Noncollision means that $X_i(t)\ne X_j(t)$ for every pair and every $t$, with probability one for each $N$. Assume that
\begin{equation}
\phi(r):=\sup_{|z|\ge r}|F(z)|<\infty\quad(r>0),
\qquad \int_1^\infty r^{d-1}\phi(r)\,\dd r<\infty.
\label{eq:SM-singular-tail}
\end{equation}
The first condition bounds the force only away from the origin and permits $\phi(r)\to\infty$ as $r\downarrow0$. The second is the \textbf{short-range} condition, including every finite-range force, which requires the force to decay faster than $r^{-d}$. For fixed $R_0>0$ and integer $M\ge1$, suppose the initial positions $x_i=X_i(0)$ satisfy
\begin{equation}
\sup_i\#\{j:|x_j-x_i|\le R_0\}\le M,
\qquad \inf_i\dist(x_i,\Gamma)\ge\ell_0>0,
\label{eq:SM-singular-density}
\end{equation}
for each $N$. $\#$ counts elements, including $i$ itself. The target $\Gamma$ may be any closed set. There are constants $c_*>0$ and $t_*>0$, independent of $N$ and of the minimum pair separation, such that
\begin{equation}
\Prob(T_N\le t)\le 2dN\exp(-c_*/t),\qquad 0<t\le t_*.
\label{eq:SM-singular-prob}
\end{equation}
Consequently (apply Markov's inequality at $t=c/\ln N$ with $c<c_*$, let $N\to\infty$, then $c\uparrow c_*$),
\begin{equation}
\liminf_{N\to\infty}(\ln N)\E T_N\ge c_*>0.
\label{eq:SM-singular-log}
\end{equation}
No $N$-independent lower bound on admissible initial pair separations or sign condition on the reciprocal force is required.
\end{theorem}

\begin{proof}
We construct the required partition by repeatedly merging nearby groups. The initial local-count bound limits the size of every group, while termination of the merging procedure guarantees separation between the groups. Put
\begin{equation}
C=16\sqrt d,\qquad
r_*=\frac{\min\{R_0,\ell_0\}}{100\sqrt d\,(C+2)^{2M}},
\qquad a=\frac{r_*}{M}.
\label{eq:SM-singular-constants}
\end{equation}
The numerical factors are chosen conservatively so that (i) a group with more than $M$ members would fit inside a ball of radius $R_0$, (ii) groups are far apart compared with their sizes, and (iii) all displacements allowed below stay smaller than $\ell_0$.
Starting from singletons, merge two groups whenever their minimum separation is at most $C$ times the largest of their two diameters and $r_*$. Here $A$ and $B$ are groups of particle labels, $|A|$ counts the labels in $A$, $d_A=\operatorname{diam}A$ is the largest initial separation within $A$, and $\dist(A,B)$ is the smallest initial separation between the two groups. A group of $m$ labels formed this way has diameter at most $r_*(C+2)^{m-1}$: by the merging rule, a merger obeys $d_{A\cup B}\le d_A+d_B+\dist(A,B)\le(C+2)\max\{d_A,d_B,r_*\}$, and induction on the number of labels gives the bound. The first group containing more than $M$ labels would contain at most $2M$ and have diameter less than $R_0$, contradicting \eqref{eq:SM-singular-density}. Thus the final partition consists of groups $A$ with
\begin{equation}
|A|\le M,\qquad
d_A:=\operatorname{diam}A\le r_*(C+2)^{M-1},\qquad
\dist(A,B)>C\max\{d_A,d_B,r_*\}.
\label{eq:SM-singular-groups}
\end{equation}
Physically, each group is a tight cluster that may contain arbitrarily close pairs, and different clusters are well separated relative to their own sizes. The separation follows from the definition of the merging, not from any assumption.

Stop if a particle in group $A$ moves a distance
$\delta_A=2\sqrt d\,(d_A+r_*)$ from its initial position (a group-dependent version of the stopping time $\sigma_\delta$ of Sec.~\ref{sec:soft}). Before this stopping time, particles in different groups remain at least half their initial distance apart, since
\begin{equation}
\delta_A+\delta_B
\le8\sqrt d\,\max\{d_A,d_B,r_*\}
<\tfrac12\dist(A,B).
\label{eq:SM-singular-separation}
\end{equation}
The total force (from other groups) on any label is therefore bounded by an $N$-independent constant $B_{\rm ext}$, similar to the force envelope~\eqref{eq:SM-BNdelta} of Sec.~\ref{sec:soft}. To see this explicitly, a ball of radius $s$ can be covered by $K_d(1+s/R_0)^d$ sets of diameter at most $R_0$, each containing at most $M$ initial particles. Count external particles by their initial distance: those within
$R_0$, and those in shells
$2^kR_0\le |x_j-x_i|<2^{k+1}R_0$.
The inner ball contains $O(M)$ particles, each exerting a force
at most $\phi(Cr_*/2)$. The $k$th shell contains $O(M2^{kd})$
particles, each exerting a force at most $\phi(2^{k-1}R_0)$,
since separations remain at least half their initial values.
Summing these contributions gives
\begin{equation}
B_{\rm ext}\le K_d M\left[
\phi(Cr_*/2)+\sum_{k=0}^{\infty}2^{kd}
\phi(2^{k-1}R_0)\right]<\infty.
\label{eq:SM-singular-external}
\end{equation}
The convergence follows from \eqref{eq:SM-singular-tail} and the monotonicity of $\phi$.

It remains to control forces inside a group without using a minimum separation. Fix a coordinate direction $e$, also allowing its negative, and write
$y_i(s)=e\cdot[X_i(s)-\sqrt{2D}W_i(s)]$ for $i\in A$. After subtracting the Brownian term, $y_i(s)$ is its initial value plus the time-integrated drift, so these functions are absolutely continuous. A group average cancels internal forces but does not control
its most advanced member. We thus average over a leading
subset, including nearby particles so that their possibly large mutual forces cancel. The following quantity implements this
choice while bounding the largest coordinate:
\begin{equation}
H_a(y)=\max_{\varnothing\ne S\subseteq A}
\left\{\frac{1}{|S|}\sum_{i\in S}y_i
+\frac{(|S|-1)a}{2}\right\}.
\label{eq:SM-singular-envelope}
\end{equation}
For a maximizing subset $S$ of size $k<|A|$, let $\mu$ be its mean. Comparing $S$ with $S\cup\{j\}$ gives
$y_j\le\mu-(k+1)a/2$ for every $j\notin S$. If $k>1$, comparison with $S\setminus\{i\}$ gives
$y_i\ge\mu-(k-1)a/2$. For $k=1$ one has $y_i=\mu$ directly. Hence
\begin{equation}
y_i-y_j\ge a,\qquad i\in S,\quad j\in A\setminus S.
\label{eq:SM-singular-active-gap}
\end{equation}
The term $(|S|-1)a/2$ favors including nearby particles.
Maximization then guarantees the gap
\eqref{eq:SM-singular-active-gap} between the selected subset
and the remaining members. Internal pair forces cancel in
the subset average, while the gap will bound cross forces
when Brownian displacements are small.
Since singletons are allowed, $H_a\ge\max_{i\in A}y_i$.

On the event
\begin{equation}
G_t=\left\{\max_{i,\alpha}\sup_{s\le t}
|\sqrt{2D}W_i^\alpha(s)|\le a/4\right\},
\label{eq:SM-singular-noise}
\end{equation}
where $\alpha=1,\ldots,d$ labels Cartesian coordinates, the physical separation of every pair crossing such an active subset is at least $a/2$, because $e\cdot(X_i-X_j)$ differs from $y_i-y_j$ by at most $a/2$. Its force is bounded by $\phi(a/2)$. All pairs within $S$ cancel in the derivative of its mean by reciprocity. If $S=A$, every internal pair cancels. A finite maximum of absolutely continuous functions is absolutely continuous, and its derivative agrees almost everywhere with that of a maximizing branch. The derivative of \eqref{eq:SM-singular-envelope} therefore obeys, almost everywhere before the stopping time and absorption,
\begin{equation}
\frac{\dd H_a}{\dd s}\le B,
\qquad B=B_0+B_{\rm ext}+(M-1)\phi(a/2).
\label{eq:SM-singular-envelope-drift}
\end{equation}
Since $y_i\le H_a(y)$ and
$H_a(y(0))\le\max_{j\in A}e\cdot x_j+(M-1)a/2$,
applying the same estimate in both coordinate directions yields
\begin{equation}
|X_i^\alpha(s)-x_i^\alpha|
\le d_A+\frac{r_*}{2}+Bs+\frac{a}{4}
\le d_A+r_*,\qquad s\le t_*:=\frac{r_*}{4B}.
\label{eq:SM-singular-close}
\end{equation}
When $B=0$, take $t_*=\infty$. Note that $B$ contains $\phi(a/2)$, which is very large for a stiff core, so $t_*$ may be very small. Both $t_*$ and the constant $c_*$ below are independent of $N$
and the smallest initial pair distance. Thus, for fixed $c>0$,
the logarithmic time $c/\ln N$ eventually becomes smaller
than $t_*$. The resulting Euclidean displacement is at most $\delta_A/2$, so continuity excludes the displacement stopping time. Moreover, \eqref{eq:SM-singular-constants}--\eqref{eq:SM-singular-groups} give $\sup_A\delta_A<\ell_0$, excluding absorption on $G_t$ for $t\le t_*$.

In the notation of Eq.~\eqref{eq:SM-semimart}, with $A_i^N(s)=\int_0^sb_i^N\,\dd r$ ($b_i^N$ being the drift in Eq.~\eqref{eq:SM-singular-model}) and $M_i^N=\sqrt{2D}W_i$, this is the setting of End Matter A. Dropping the Brownian term $a/4$ in Eq.~\eqref{eq:SM-singular-close} gives $|A_i^N(s)|\le\delta_A/2<\ell_0/2$, so for $t\le t_*$ the noise event $G_t$ lies inside the good event $G_{N,t}(\ell_0/2)$ of Eq.~\eqref{eq:SM-buffered-master}. This displacement budget replaces the force budget of Theorem~1. On $G_t$, we also have
$|M_i^N(s)|\le\sqrt d\,a/4<\ell_0/2$.
The total displacement is therefore smaller than $\ell_0$,
so no particle can reach the target on $G_t$.
Hence $\Prob(T_N\le t)\le\Prob(G_t^c)$. The reflection principle and a union bound over labels and coordinates now give
\begin{equation}
\Prob(T_N\le t)\le\Prob(G_t^c)
\le2dN\,\erfc\!\left(\frac{a}{4\sqrt{4Dt}}\right)
\le2dN\exp\!\left(-\frac{a^2}{64Dt}\right).
\end{equation}
Thus $c_*=a^2/(64D)$ is admissible. On the logarithmic window $t=c/\ln N$, with $0<c<c_*$, the good-event failure probability is at most $2dN^{1-c_*/c}\to0$. Markov's inequality $\E T_N\ge t\Prob(T_N>t)$ then proves \eqref{eq:SM-singular-log}.
\end{proof}

For the LJ potential
$U(r)=4\varepsilon[(\sigma/r)^{12}-(\sigma/r)^6]$, where $\varepsilon>0$ and $\sigma>0$ set the energy and length scales,
the reciprocal force $F(z)=-\nabla U(|z|)$ is bounded away from $z=0$ and decays as $r^{-7}$. The theorem therefore covers the full potential, including attraction, in every physical dimension $d=1,2,3$ (more generally, $d<7$). The WCA potential is $U(r)+\varepsilon$ for $r<2^{1/6}\sigma$ and zero beyond that distance, so its force has finite range and also satisfies the theorem. For standard finite-$N$ LJ dynamics, global existence and noncollision from distinct deterministic initial points follow from the singular repulsive core \cite{MoraleRuiUgolini2026}. The difference between the WCA and LJ forces is bounded, so Girsanov's theorem \cite{KaratzasShreve} transfers the probability-zero collision event to WCA on each finite interval. Bounded additional drift is handled in the same way. Thus noncollision imposes no extra initial-spacing condition for these two models. The potential parameters, $R_0$, and $M$ remain fixed as $N$ grows. This proves case (ii) of Proposition~1. The explicit constant above is conservative: this proof establishes the logarithmic scale, without asserting the sharp coefficient $\ell_0^2/(4D)$. %Combined with a matching upper bound, for example the flat-target comparison for WCA repulsion in Sec.~\ref{sec:upper-product}, it gives $\E T_N=\Theta((\ln N)^{-1})$ when at least $N^\gamma$ particles, for some fixed $\gamma>0$, start within a fixed target distance; indeed, $\limsup_N(\ln m_N)\E T_N\le\ell_+^2/(4D)$ with $\ln m_N\ge\gamma\ln N$.

\section{Proof of Proposition 2: upper bounds for the mean EFPT}
\label{sec:upper}
Here, we prove the logarithmic upper bounds for the mean EFPT in Proposition~2 of the main text. Sections~\ref{sec:upper-window} and~\ref{sec:upper-product} prove its settings (i) and (ii), respectively.

\subsection{Setting (i): a EFPT probability bound and confinement}
\label{sec:upper-window}

This subsection proves setting (i) of Proposition~2. We first show that, with probability tending to one, the EFPT is at most of order $1/\ln m_N$. Consider the flat target $z=0$, with searchers on $z>0$. Here $z_i$ is particle $i$'s normal coordinate, positive away from the target, and $Z=(z_1,\ldots,z_N)$ collects these coordinates. Let $I_N$ contain $m_N\to\infty$ candidate labels with $z_i(0)\le\ell_+$, where $\ell_+$ is a $N$-independent upper bound on their initial target distances. Up to $T_N$, suppose
\begin{equation}
z_i(s)=z_i(0)+V_i(s)+\sqrt{2D}\,\beta_i(s),\qquad i\in I_N,
\end{equation}
where the $\beta_i$ are independent one-dimensional Brownian motions and $V_i$ is the remaining normal displacement, positive away from the target. For a displacement allowance $\rho\ge0$, put
$G_N(t,\rho)=\{\max_{i\in I_N}\sup_{s\le t\wedge T_N}V_i(s)\le\rho\}$.
If a Brownian driver crosses $-(\ell_++\rho)/\sqrt{2D}$ while this event holds, the interacting system must already have hit. Thus
\begin{equation}
\Prob(T_N>t)\le\Prob(G_N(t,\rho)^c)+
\left[1-\erfc\left(\frac{\ell_++\rho}{\sqrt{4Dt}}\right)\right]^{m_N}.
\label{eq:SM-oneshot-upper-bound}
\end{equation}
The product concerns independent driving events, and no factorization of the interacting hitting times is assumed. Indeed, on $G_N(t,\rho)\cap\{T_N>t\}$ every candidate keeps $z_i(s)>0$, hence $\sqrt{2D}\beta_i(s)>-z_i(0)-V_i(s)\ge-(\ell_++\rho)$ for all $s\le t$. The probability that all $m_N$ independent drivers stay above this level is the product in Eq.~\eqref{eq:SM-oneshot-upper-bound}. Interactions may correlate the positions arbitrarily, but not the thermal noises.

At $t_N=c/\ln m_N$, a deterministic outward speed bound $B_N=o(\ln m_N)$ gives $\rho=B_Nt_N=o(1)$. For every $c>\ell_+^2/(4D)$, Eq.~\eqref{eq:SM-oneshot-upper-bound} then bounds survival by $\exp[-m_N^\epsilon]$ for some $\epsilon>0$, since each candidate crosses with probability $m_N^{-\ell_+^2/(4Dc)+o(1)}$ and $(1-p)^{m}\le e^{-mp}$. A bound $B_N\le\alpha\ln m_N$ gives the same conclusion whenever $4Dc>(\ell_++\alpha c)^2$.  
Such a $c$ exists if $\alpha\ell_+<D$. This is the upper-bound counterpart of the condition above Eq.~\eqref{eq:SM-calpha}: an adverse drift lengthens the distance the noise must cover from $\ell_+$ to $\ell_++\alpha c$.

A high probability of an early hit does not yet bound the mean. Rare realizations in which no particle arrives within the window may last very long or never end. For example, independent searchers with a constant outward drift hit the target within the logarithmic window with probability tending to one, yet with positive probability none of them ever arrives, so $\E T_N=\infty$. A bound on the mean therefore also requires control of these rare long-time realizations. The two settings of Proposition~2 provide it in two ways. In setting (i), confinement bounds the remaining mean search time from every surviving configuration, as shown next. In setting (ii), the survival-product comparison of Sec.~\ref{sec:upper-product} bounds the survival probability at all times, which controls the early hit and the long-time tail at once.

To control the mean, let $R_N(t)$ be a deterministic upper bound on the conditional remaining mean time, on $\{T_N>t\}$, i.e., $\E[T_N-t\,|\,\mathcal F_t]\le R_N(t)$ there, where $\mathcal F_t$ is the history up to time $t$. The identity for conditional expectations yields
\begin{equation}
\E T_N\le t+R_N(t)\Prob(T_N>t).
\label{eq:SM-residual}
\end{equation}
Consequently a first-window bound $\Prob(T_N>t_N)\le\varepsilon_N$ gives $\E T_N\le t_N+o(1/\ln m_N)$ whenever $R_N(t_N)\varepsilon_N=o(1/\ln m_N)$.

Confinement justifies this tail condition. Suppose $0<z_i<L$, where $L>0$ is a fixed slab width, with reflection or another non-outward push at $L$, and the total \emph{outward} drift of each candidate is at most $B_N\ge0$. The mean hitting time $u(z)$ for the comparison diffusion with constant drift $B_N$, reflected at $L$, solves $Du''+B_Nu'=-1$, $u(0)=0$, $u'(L)=0$. For $B_N>0$, solving the equation gives
$u'(z)=[e^{B_N(L-z)/D}-1]/B_N\ge0$.
Writing $b_i^Z(s)\le B_N$ for the actual normal drift,
the It\^o drift of $u(z_i(s))+s$ is
$Du''+b_i^Zu'+1=(b_i^Z-B_N)u'\le0$.
Reflection at $L$ contributes nothing because $u'(L)=0$,
and inward pushes contribute nonpositively.
Thus $u(z_i(s))+s$, stopped at $T_N$, is a supermartingale.
Taking expectations and using $u\ge0$ gives (supermartingale property)
\begin{equation}
\E[u(z_i(t\wedge T_N))+t\wedge T_N]\leq u(z_i(0))\Rightarrow\E(T_N\wedge t)
\le u(z_i(0))-\E u(z_i(T_N\wedge t))
\le u(L).
\end{equation}
Letting $t\to\infty$ yields $\E T_N\le u(L)$.
The same argument applies conditionally at any time before
the first hit, so the remaining mean time $R_N$ is also
bounded by $u(L)$.
Integrating $u'$ gives $u(L)=C(B_N,L,D)$ below,
with the $B_N=0$ case obtained by continuity:
\begin{equation}
R_N\le C(B_N,L,D),\qquad
C(B,L,D)=\frac D{B^2}(e^{BL/D}-1)-\frac LB,
\quad C(0,L,D)=\frac{L^2}{2D}.
\label{eq:SM-C-BLD}
\end{equation}
For $B_N=O(\ln m_N)$ this grows at most polynomially in $m_N$, since $C(B_N,L,D)\le(D/B_N^2)e^{B_NL/D}$ and $e^{B_NL/D}=m_N^{\alpha L/D}$ for $B_N=\alpha\ln m_N$, so the preceding exponentially small survival probability controls the residual. This proves setting (i) of Proposition~2 of the main text and, when $B_N=o(\ln m_N)$,
{
\begin{equation}
\limsup_{N\to\infty}(\ln m_N)\E T_N\le\frac{\ell_+^2}{4D}.
\label{eq:SM-confined-sharp}
\end{equation}
}
The sufficient restriction $\alpha L<D$ in the main text implies $\alpha\ell_+<D$. 

\subsection{Setting (ii): pure repulsion}
\label{sec:upper-product}

This subsection proves setting (ii) of Proposition~2. Repulsion pushes one member of a pair away from a target and the other toward it. We show that this cannot delay the first arrival relative to independent Brownian searchers at any time by controlling the early hit and the long-time tail at once. In a word, the push toward the target wins because survival is more sensitive to the particle that is closer to the target.

Let $\operatorname{erf}=1-\erfc$ denote the error function. Define the single-particle Brownian survival probability on the half-line and its logarithmic derivative by
\begin{equation}
q(z,t)=\operatorname{erf}\left(\frac z{\sqrt{4Dt}}\right),\qquad
h(z,t)=\partial_z\ln q(z,t).
\end{equation}
For $z,t>0$, $h>0$ and
\begin{equation}
\partial_z h=-\frac z{2Dt}h-h^2<0.
\label{eq:SM-h-monotone}
\end{equation}
Here $q(z,t)$ is the probability that a free Brownian searcher started at distance $z$ has not reached the target by time $t$, and $h$ measures how sensitive this survival is to the distance. Equation~\eqref{eq:SM-h-monotone} states that the sensitivity decreases with $z$: a small displacement matters more for a particle close to the target.
The product $U(Z,t)=\prod_iq(z_i,t)$ obeys
\begin{equation}
(\partial_t-\mathcal L)U=-U\sum_i b_i^Z h(z_i,t).
\label{eq:SM-U-supersolution}
\end{equation}
Here $\mathcal L$ is the full backward generator, and $b_i^Z$ is the drift in the $z_i$ direction. It may depend on tangential or internal variables. The product $U$ is the survival probability the system would have if all particles were independent free Brownian searchers. Equation~\eqref{eq:SM-U-supersolution} follows from the heat equation $\partial_tq=D\partial_z^2q$, so the drift enters only through the rate $\sum_ib_i^Z\partial_{z_i}\ln U$. 

Thus, assuming $\sum_i b_i^Zh(z_i,t)\le0$ makes $U$ a supersolution, which is an upper comparison function for the interacting survival probability. To see this, note that the true survival probability $Q$ for the interacting system solves $\partial_tQ_N=\mathcal LQ_N$, whereas $U$ satisfies $\partial_tU\ge\mathcal LU$ with the same initial and absorbing boundary values, so $U$ stays above $Q_N$ by the parabolic comparison principle \cite{Lieberman1996}: at a spatial minimum of $U-Q_N$ the gradient vanishes and the Laplacian is nonnegative, so $\partial_t(U-Q_N)\ge\mathcal L(U-Q_N)\ge0$ there, and a negative minimum cannot form. Consequently,
\begin{equation}
\Prob(T_N>t)\le\prod_iq(z_i(0),t).
\label{eq:SM-flat-survival-domination}
\end{equation}
At a reflecting wall $z_i=L$, the survival probability satisfies
$\partial_{z_i}Q_N=0$, while $\partial_{z_i}U\ge0$.
Thus $U$ also satisfies the boundary inequality required
by the comparison principle, and the same bound holds
with or without this confinement.
%At a reflecting wall $z_i=L$, the additional term in It\^o's formula is $-\partial_{z_i}U\,\dd K_i^L\le0$, since $\partial_{z_i}U\ge0$ and the reflection regulator $K_i^L$ is nondecreasing.Thus Eq.~\eqref{eq:SM-flat-survival-domination} and theresulting mean-time bound hold with or without this confinement.

We now verify the condition $\sum_ib_i^Zh(z_i,t)\le0$ for reciprocal central repulsion with a targetward or vanishing remaining drift. For reciprocal central repulsion,
$b_i^Z=c_i+\sum_{j\ne i}a_{ij}(z_i-z_j)$ with $c_i\le0$ and $a_{ij}=a_{ji}\ge0$. Here $c_i$ is the remaining normal drift, and $a_{ij}$ is the repulsive force magnitude per unit pair separation. Pairing the forces gives
\begin{equation}
\sum_i b_i^Zh(z_i,t)=\sum_i c_i h(z_i,t)
+\sum_{i<j}a_{ij}(z_i-z_j)[h(z_i,t)-h(z_j,t)]\le0.
\label{eq:SM-pair-sign}
\end{equation}
Each pair term combines a push of the farther particle away from the target with an equal push of the closer particle toward it. Since $h$ is larger for the closer particle [Eq.~\eqref{eq:SM-h-monotone}], the net contribution of the pair to the drift term of $\dd U$ is nonpositive: repulsion can only reduce survival. No force-magnitude restriction enters this proof. A singular force can be handled by stopping away from collisions and removing that localization for a noncolliding solution. For a pair potential $V(r)$ and separation $r_{ij}=|X_i-X_j|$, $a_{ij}=-V'(r_{ij})/r_{ij}\ge0$ precisely when the force is repulsive.

If $m=m_N$ candidates satisfy $z_i(0)\le\ell_+$, Eq.~\eqref{eq:SM-flat-survival-domination} is at most $q(\ell_+,t)^m$. Its integral has upper asymptotic $\ell_+^2/(4D\ln m)$: split at $t_m=(1+\epsilon)\ell_+^2/(4D\ln m)$ and a fixed $T>\ell_+^2/(\pi D)$. On $[t_m,T]$ the integral is exponentially small in a positive power of $m$. For $m>2$,
\begin{equation}
\int_T^\infty q(\ell_+,t)^m\,\dd t
\le\frac{2T}{m-2}\left(\frac{\ell_+}{\sqrt{\pi DT}}\right)^m.
\end{equation}
For $t>T$ each factor is at most $\ell_+/\sqrt{\pi Dt}<1$, so the product decays as $t^{-m/2}$ and its integral is negligible. This is how the long-time tail is controlled without confinement. Letting $\epsilon\downarrow0$ proves setting (ii) of Proposition~2 of the main text, including control of the long-time tail.

\section{Acceleration limits and examples}
\label{sec:acceleration}

In this section, we provide additional details on Theorem 2 and Examples A-C of the main text.
%Interactions can speed up search by increasing either the directed force or the noise strength as particles are added. We bound the resulting acceleration and construct examples of both mechanisms. For continuous (diffusive) dynamics at a fixed initial gap, these are the only two interaction-driven resources: Sec.~\ref{sec:accel-master} turns this statement into a quantitative bound, and Secs.~\ref{sec:screened}--\ref{sec:coulomb} realize both mechanisms.

\subsection{More details on Theorem 2 of the main text}
\label{sec:accel-master}

%Directed motion and noise must together cover the initial target gap. Their strengths set a lower bound on the first-arrival time, even when particle trajectories are correlated. The bound interpolates between the logarithmic law with an enhanced diffusivity and ballistic crossing of the gap by the drift.
This subsection gives more details on Theorem~2 than End Matter: where the threshold equation for $t_N^\star$ comes from and what its limiting regimes mean, and how $(p+1)$-body force channels and stochastic noise channels determine $B_N$ and $D_N^{\rm eff}$, which yields the scaling branches of Table~I of the main text. For completeness, we first restate the bound proved in End Matter C.

For a convex target, project the displacement along the supporting direction $e_i$:
\begin{equation}
e_i\cdot[X_i^N(s)-x_i^N]=A_i^N(s)+M_i^N(s).
\label{eq:SM-accel-decomp}
\end{equation}
Here $A_i^N$ is the accumulated displacement in direction $e_i$ from drift and any explicit boundary pushes. It is continuous and of finite variation with $A_i^N(0)=0$. The scalar fluctuating displacement $M_i^N$ is a continuous local martingale starting at zero. They may be correlated between labels. This is the general form of the argument of Sec.~\ref{sec:projection}: $A_i^N$ collects all directed motion, and $M_i^N$ all fluctuating motion, whose strength may now depend on the state and grow with $N$. Suppose that on an event $G_{N,t}$,
$\max_i\sup_{s\le t\wedge T_N}A_i^N(s)\le u_N(t)<\ell_0$, where $u_N(t)\ge0$ is a deterministic displacement bound. For a threshold $a>0$, define
\begin{equation}
\Psi_N(t,a)=\max_i\Prob\left(G_{N,t}\cap
\left\{\sup_{s\le t\wedge T_N}M_i^N(s)\ge a\right\}\right).
\end{equation}
The first-hitter argument of Sec.~\ref{sec:projection} gives
\begin{align}
\Prob(T_N\le t)&\le\Prob(G_{N,t}^c)+N\Psi_N(t,\ell_0-u_N(t)),\\
\E T_N&\ge t[1-\Prob(G_{N,t}^c)-N\Psi_N(t,\ell_0-u_N(t))]_+.
\label{eq:SM-unified-mean}
\end{align}
Indeed, a first hitter must supply the distance left after its finite-variation displacement through its own martingale excursion.

Write $\langle M_i^N\rangle_s$ for the quadratic variation accumulated up to time $s$. If $u_N(t)=B_Nt$ and $\langle M_i^N\rangle_s\le2D_N^{\rm eff}s$ on $G_{N,t}$, with deterministic $B_N\ge0$ and $D_N^{\rm eff}>0$ bounding the drift and effective diffusivity, the Brownian time-change representation \cite{RevuzYor} bounds $\Psi_N(t,a)$ by $\erfc(a/\sqrt{4D_N^{\rm eff}t})$. With this bound, Eq.~\eqref{eq:SM-unified-mean} gives $\E T_N\ge t[1-o(1)]$ as long as $\Prob(G_{N,t}^c)\to0$ and $N\erfc[(\ell_0-B_Nt)/\sqrt{4D_N^{\rm eff}t}]\to0$. Since $\erfc(x)\le e^{-x^2}/(\sqrt\pi x)$, the latter holds when $(\ell_0-B_Nt)^2/(4D_N^{\rm eff}t)\ge\ln N$, so the largest admissible time is the one at which equality holds. Taking the square root, this threshold time solves
\begin{equation}
B_Nt+2\sqrt{D_N^{\rm eff}t\ln N}=\ell_0,
\label{eq:SM-threshold}
\end{equation}
whose positive solution is
\begin{equation}
t_N^\star=
\frac{\ell_0^2}{\bigl(\sqrt{D_N^{\rm eff}\ln N+B_N\ell_0}
+\sqrt{D_N^{\rm eff}\ln N}\bigr)^2}.
\label{eq:SM-tstar}
\end{equation}
Equation~\eqref{eq:SM-threshold} has the same structure as the condition above Eq.~\eqref{eq:SM-calpha}: $t_N^\star$ is the time at which the directed displacement $B_Nt$ plus the largest noise excursion among $N$ searchers, $\sqrt{4D_N^{\rm eff}t\ln N}$, just equals the gap. Two limits are instructive. If $B_N\ell_0\ll D_N^{\rm eff}\ln N$, then $t_N^\star\simeq\ell_0^2/(4D_N^{\rm eff}\ln N)$, the logarithmic law with an enhanced diffusivity. If $B_N\ell_0\gg D_N^{\rm eff}\ln N$, then $t_N^\star\simeq\ell_0/B_N$, the time needed by the drift alone. The intermediate case $D_N^{\rm eff}=D$, $B_N=\alpha\ln N$ gives $t_N^\star=c_-(\alpha)/\ln N$, i.e., Eq.~\eqref{eq:SM-calpha}.
At this time, $N\erfc(\sqrt{\ln N})\to0$. Indeed, $\erfc(y)\le e^{-y^2}/(\sqrt\pi y)$ bounds it by $(\pi\ln N)^{-1/2}$. Hence $\Prob(G_{N,t_N^\star}^c)\to0$ implies $\E T_N\ge t_N^\star(1-o(1))$, proving Theorem~2 and End Matter C.

With bounded $D_N^{\rm eff}$ and $B_N\le BN^p$, where $B>0$ is independent of $N$ and $p>0$, this gives
$\liminf N^p\E T_N\ge\ell_0/B$. A label belongs to at most
$\binom{N-1}{p}$ deterministic $(p+1)$-particle channels, i.e., interacting $(p+1)$-tuples containing that label, so uniformly bounded force per channel gives the $N^{-p}$ scale for integer $p$. For bounded drift and $D_N^{\rm eff}=O(N^q)$ with $q\ge0$, as produced by $O(N^q)$ uncorrelated zero-mean noise channels of bounded variance, the lower scale is $(N^q\ln N)^{-1}$. %In a sum of stochastic channels, variances add only for orthogonal martingale increments; a common noise can add amplitudes coherently and must be assessed through its actual quadratic variation. For example, $K$ channels driven by one shared noise produce a variance proportional to $K^2$ rather than $K$.

\subsection{Example A: screened single-file repulsion}
\label{sec:screened}

%In a channel where particles cannot pass one another, screened repulsion pushes the leading particle toward the absorbing end. %Adding particles at fixed channel length increases this push; contact reflection also acts toward the target. 
Here we provide more details of the Example A of the main text, including a tighter upper bound and a matching lower bound. Recall that this example deliberately violates the initial-density condition of Proposition~1: with $N$ particles in a fixed length $L$, every particle has $N-1$ neighbors within distance $L$.

Consider ordered point particles $0\le X_1\le\cdots\le X_N\le L$, with no-passing reflection at contacts, reflection at the left wall, and absorption when $X_N=L$. The particles interact through the potential
\begin{equation}
U_N(x)=g\sum_{i<j}e^{-(x_j-x_i)/\lambda},\qquad g,\lambda>0,
\end{equation}
where $g$ sets the interaction strength and $\lambda$ is the screening length. The potential gives the leader a drift at least
$v_N=(g/\lambda)e^{-L/\lambda}(N-1)$, because each of the $N-1$ particles behind the leader lies within distance $L$ and pushes it forward with force at least $(g/\lambda)e^{-L/\lambda}$. Its collision local times also push targetward. Stopping at $t\wedge T_N$ and taking expectations gives
$\E X_N(t\wedge T_N)\ge X_N(0)+v_N\E(t\wedge T_N)$,
because the stopped Brownian term has zero mean and
the reflection terms are nonnegative.
Since $X_N(t\wedge T_N)\le L$, letting $t\to\infty$ yields
\begin{equation}
\E T_N\le \frac{L-X_N(0)}{v_N}
=\frac{\lambda e^{L/\lambda}}{g}
\frac{L-X_N(0)}{N-1},
\end{equation}
which is the formula presented in the main text. Notably, the comparison principle can yield a tighter upper bound. The idea is to compare the leader with a single particle that feels only this weakest possible push. Since the leader is always pushed at least as hard, it stays ahead of the comparison particle. Couple the leader to $Y$, started at $Y(0)=X_N(0)$ and driven by the same Brownian motion, with constant drift $v_N$ and reflection at zero. Tanaka's formula \cite{KaratzasShreve} for $(Y-X_N)^+$ proves $Y\le X_N$: on $\{Y>X_N\}$ the reflection of $Y$ vanishes, the drift difference is nonpositive, and the collision terms of $X_N$ enter with a negative sign. The Brownian parts cancel. The comparison mean solves $Du''+v_Nu'=-1$, $u'(0)=0$, $u(L)=0$, giving
\begin{equation}
\E T_N\le u(X_N(0))
=\frac{L-X_N(0)}{v_N}
-\frac D{v_N^2}\left(e^{-v_NX_N(0)/D}-e^{-v_NL/D}\right)
\le\frac{\lambda e^{L/\lambda}}g\,\frac{L-X_N(0)}{N-1}.
\end{equation}

\emph{Lower bound.} Theorem~2 cannot be applied directly to obtain the lower bound due to the no-passing reflection. %these push the leader toward the target and are not controlled by a force budget. 
Fortunately, a no-passing reflection between identical particles is equivalent to simply exchanging their labels \cite{Harris1965}. Hence $T_N$ has the same law as for particles that may pass one another, interact through $g\sum_{i<j}e^{-|x_j-x_i|/\lambda}$, and are reflected at $x=0$. Each of their drifts is bounded by $B_N=(g/\lambda)(N-1)$. Their only other finite-variation term, the wall push, is at most $B_Nt+\sup_{s\le t}|\sqrt{2D}W_i(s)|$ up to time $t$, by the Skorokhod map. On the event $\max_i\sup_{s\le t}|\sqrt{2D}W_i(s)|\le B_Nt$, every targetward finite-variation displacement is therefore at most $3B_Nt$, and this event fails with probability at most $2N\erfc(B_N\sqrt{t/4D})$. Theorem~2 thus applies with drift budget $3B_N$ and $D_N^{\rm eff}=D$. At its time $t_N^\star\simeq\ell_0/(3B_N)$, where $\ell_0=L-X_N(0)$ is the initial gap of the leader, this failure probability vanishes. Combined with the upper bound above,
\begin{equation}
\frac{\lambda\ell_0}{3g}\le\liminf_{N\to\infty}N\E T_N\le\limsup_{N\to\infty}N\E T_N\le\frac{\lambda e^{L/\lambda}\ell_0}{g},
\end{equation}
so $\E T_N=\Theta(N^{-1})$.

%This proves Example A. Its reflection terms are explicit; a smooth realization of the deterministic channel bound is given in Sec.~\ref{sec:saturation}.

\subsection{Example B: antisymmetric stochastic pair forcing}
\label{sec:kicks}

In this example, each particle pair receives opposite random kicks, so every particle collects noise from $N-1$ partners. We show how the resulting enhanced diffusivity accelerates the first arrival despite correlations between particles. %The correlations turn out to be weak: they amount to a single common displacement with an $N$-independent diffusivity, negligible compared with the individual diffusivity $\simeq\kappa N$.

On $[0,L]$, reflected at zero and absorbed at $L$, start all labels at $x_0=L-\ell_0\ge0$ and consider
{
\begin{equation}
\dd X_i=\sqrt{2D}\,\dd W_i+
\sqrt{2\kappa}\left(\sum_{j<i}\dd B_{ji}-\sum_{j>i}\dd B_{ij}\right)
+\dd K_i^0.
\label{eq:SM-kick-model}
\end{equation}
}Here $\kappa>0$ is the pair-noise strength. All $W_i$ and $B_{ij}$ ($i<j$) are independent standard Brownian motions, and $K_i^0$ is the nondecreasing displacement imposed by reflection at zero. The driver of each coordinate has diffusivity $D_N=D+\kappa(N-1)$, because the $N-1$ pair noises acting on one particle are independent, so their variances add. Denote $\tau_i$ as the particle $i$'s first arrival at $L$. For each particle separately, Brownian motion reflected at $0$
has the same law as the absolute value of Brownian motion on
the full line, with the same initial position and diffusivity \cite{RevuzYor}. This gives
\begin{equation}
\Prob(\tau_i\le t)\le
\erfc\left(\frac{\ell_0}{\sqrt{4D_Nt}}\right)
+\erfc\left(\frac{L+x_0}{\sqrt{4D_Nt}}\right).
\end{equation}
Indeed, the reflected coordinate has the law of $|x_0+M_i|$, with $M_i$ its unreflected driver, and reaches $L$ only if $M_i$ reaches $\ell_0=L-x_0$ or $-(L+x_0)$. The two possible image crossings are bounded by a union bound. The image bound above differs from the zero-drift estimate
of Theorem~2 by at most a factor of two. With
$D_N^{\rm eff}=D_N\sim\kappa N$, it therefore yields
$\liminf_{N\to\infty}N\ln N\,\E T_N\ge\ell_0^2/(4\kappa)$.

For the upper bound, the vector of pair drivers has the same Gaussian law as
$\sqrt N(B_i-\overline B)_{i=1}^N$, where $B_1,\ldots,B_N$ are independent standard Brownian motions, also independent of the $W_i$, and $\overline B=N^{-1}\sum_iB_i$.
Thus the unreflected coordinate drivers can be written as
\begin{equation}
M_i=Z_i-C,\qquad
Z_i=\sqrt{2D}W_i+\sqrt{2\kappa N}B_i,\qquad
C=\sqrt{2\kappa N}\,\overline B.
\end{equation}
Here $M_i$, $Z_i$, and $C$ are displacement processes starting at zero. The $Z_i$ are independent with diffusivity $D+\kappa N$, while $C$ has diffusivity $\kappa$. They need not be independent of $C$. This representation carries all correlations among particles in the single common displacement $C$, whose diffusivity $\kappa$ does not grow with $N$, whereas each $Z_i$ has diffusivity $\simeq\kappa N$. On the time scale $t\sim1/(N\ln N)$ the common part is negligible, and the $Z_i$ act as $N$ independent searchers with diffusivity $\simeq\kappa N$. This explains the scale $\ell_0^2/(4\kappa N\ln N)$. On $\{T_N>t\}$, nonnegative reflection implies $M_i(t)<\ell_0$ for all $i$. Hence, for every $\epsilon>0$,
\begin{equation}
\Prob(T_N>t)\le
2e^{-\epsilon^2/(4\kappa t)}
+e^{-Np_N(t)},\qquad
p_N(t)=\tfrac12\erfc\left(\frac{\ell_0+\epsilon}{\sqrt{4(D+\kappa N)t}}\right).
\label{eq:SM-kick-survival-upper}
\end{equation}
Indeed, on $\{T_N>t\}$ either $|C(t)|>\epsilon$, or all $Z_i(t)<\ell_0+\epsilon$. By independence of the $Z_i$, the latter has probability $(1-p_N)^N\le e^{-Np_N}$. Take $t=c/(N\ln N)$ with $c>\ell_0^2/(4\kappa)$, and choose $\epsilon$ so that $(\ell_0+\epsilon)^2<4\kappa c$. Both terms vanish faster than any inverse power of $N$. Given the surviving configuration, a selected label has remaining mean at most $L^2/(2D_N)$, since its own coordinate is a reflected Brownian motion with diffusivity $D_N$ on $[0,L]$. Equation~\eqref{eq:SM-residual} then yields
\begin{equation}
\lim_{N\to\infty}N\ln N\,\E T_N=\frac{\ell_0^2}{4\kappa}.
\label{eq:SM-kick-exact}
\end{equation}

%This proves Example B. If $x_0>0$, the second image term has a strictly larger short-time exponent; for $x_0=0$ both terms have the same exponent, leaving the leading constant unchanged.

\subsection{An additional example that can attain the many-body interaction acceleration limit}
\label{sec:saturation}

The construction in this section attains the many-body acceleration limit in the main text (Theorem 2). It is a proof of principle that the $O(N^{-p})$ bound there (see also Sec.~\ref{sec:accel-master}) cannot be improved in general. %rather than a model of a specific physical system.

Consider one-dimensional motion without reflecting walls, with the same thermal diffusivity $D$. Fix an integer $r\ge2$, the number of particles in each interaction. Let every coordinate initially equal zero and place the target at $L=\ell_0$. Choose a smooth nonlinear function $\chi$ such that $1/2\le\chi'\le1$ everywhere and $\chi'=1$ on $[-3\ell_0,3\ell_0]$. For each $(r-1)$-element subset $S$ of the other labels, couple it to the tagged label $N$ through
\begin{equation}
V_S(x)=-f\,\chi\left(x_N-\frac1{r-1}\sum_{j\in S}x_j\right),\qquad f>0.
\end{equation}
The total potential is $\sum_SV_S$, and $f$ sets the force scale of each interaction. These potentials are smooth, translation invariant, and exert zero total force. Set $v_N=f\binom{N-1}{r-1}$. The tagged drift lies between $v_N/2$ and $v_N$, and each other drift lies between $-v_N/(N-1)$ and $-v_N/[2(N-1)]$. Thus every label has drift magnitude at most $v_N$, and the acceleration limit (Theorem 2) gives
$\liminf v_N\E T_N\ge\ell_0$, since Eq.~\eqref{eq:SM-tstar} with $B_N=v_N\gg\ln N$ and $D_N^{\rm eff}=D$ gives $t_N^\star\simeq\ell_0/v_N$.

For the matching upper bound, we show that the tagged particle is pushed with the full force $v_N$ throughout the relevant window and therefore arrives within a time $\simeq\ell_0/v_N$. For fixed $0<\epsilon<1/4$, take $t_N=(1+\epsilon)\ell_0/v_N$. With probability tending to one, all Brownian displacements up to $t_N$ are at most $\ell_0/4$, since
$N\exp[-c/t_N]\to0$. The deterministic drift bounds then keep every argument of $\chi'$ inside $[-3\ell_0,3\ell_0]$ throughout this window for large $N$. The tagged drift is exactly $v_N$ there. Survival at $t_N$ then requires $\sqrt{2D}W_N(t_N)<-\epsilon\ell_0$, whose probability tends to zero. Finally, the tagged coordinate dominates a Brownian motion with drift $v_N/2$. Its hitting time $\widehat\tau_N$ has
\begin{equation}
\E\widehat\tau_N=\frac{2\ell_0}{v_N},\qquad
\E\widehat\tau_N^2=\frac{4\ell_0^2}{v_N^2}
+\frac{16D\ell_0}{v_N^3}.
\end{equation}
Since $T_N\le\widehat\tau_N$, these moments make $v_NT_N$ uniformly integrable. The high-probability upper estimate therefore also controls the mean. Letting $\epsilon\downarrow0$ yields
\begin{equation}
\lim_{N\to\infty}N^{p}\E T_N=\frac{p!\,\ell_0}{f},
\end{equation}
where $p=r-1$. Thus, this example saturates the $O(N^{-p})$ scale in the main text. %The nonlinearity outside the force plateau supplies a genuine position-dependent interaction while retaining exact asymptotic saturation. If $\chi$ were linear, all forces would be constant, and the construction would reduce to constant external forces on the individual particles.

\subsection{Example C: unscreened Coulomb repulsion with $N$-independent initial density}
\label{sec:coulomb}

A three-dimensional cloud with unscreened Coulomb repulsion can exert a growing force on its leading particle even at $O(1)$ initial number density. In this case, very far-away particles can contribute to speeding up the extreme search. This is because the Coulomb interaction considered here is long-range: its tail violates the integrability condition~\eqref{eq:SM-singular-tail}.

Consider three-dimensional particles with same-sign Coulomb repulsion,
\begin{equation}
\dd X_i=g\sum_{j\ne i}\frac{X_i-X_j}{|X_i-X_j|^3}\,\dd t
+\sqrt{2D}\,\dd W_i,\qquad g,D>0,
\label{eq:SM-coulomb-model}
\end{equation}
Here $g$ is the fixed Coulomb coupling, which is $N$-independent. The target is the plane $z=0$, with $z$ the first Cartesian coordinate. Fix a lattice spacing $a>0$ and a target gap $0<\ell_0\le a/2^{16}$. For an integer $J\ge1$, label the leading particle by $0$ and set
\begin{equation}
\begin{aligned}
x_0&=(\ell_0,0,0),\qquad
x_{mjk}=(\ell_0+am,aj,ak),\\
1&\le m\le J,\qquad -J\le j,k\le J,
\qquad N=1+J(2J+1)^2.
\end{aligned}
\label{eq:SM-coulomb-cloud}
\end{equation}
The cloud occupies the side opposite the target. Every ball of fixed radius contains a number of initial particles bounded independently of $N$, and all initial pair distances are at least $a$. Writing $b_{0z}$ for the normal component of the leader's drift, its initial targetward drift is
\begin{equation}
-b_{0z}(x)=\frac{g}{a^2}
\sum_{m=1}^{J}\sum_{j,k=-J}^{J}
\frac{m}{(m^2+j^2+k^2)^{3/2}}
=\Theta(gJ/a^2).
\label{eq:SM-coulomb-initial-force}
\end{equation}
At fixed number density, a hemispherical shell of radius $r$ and width $\dd r$ behind the leader contains $\propto r^2\dd r$ particles, each exerting a force $\propto r^{-2}$. The two factors compensate, so every shell contributes an order-one targetward push, and summing over the cloud depth $aJ$ produces a force $\propto J$. Here $\phi(r)=g/r^2$, so the tail integral in Eq.~\eqref{eq:SM-singular-tail} diverges in three dimensions.

We first bound the hitting probability. The cloud particles also repel each other and begin to spread. We show that on the short window of interest they barely move, so the push on the leader retains its order of magnitude. Stop when any particle moves by $\delta=a/8$ from its initial position, or when a particle reaches the target. Before this stopping time, every pair distance is at least 3/4 of its initial value. Counting lattice sites in successive cubic shells, each containing at most $24q^2+2$ sites at index $q$, gives
\begin{equation}
\max_i|b_i|\le B_J:=\frac{96gJ}{a^2},\qquad
-b_{0z}\ge v_J:=\frac{gJ}{16a^2}.
\label{eq:SM-coulomb-budgets}
\end{equation}
For the first bound, shell indices run to at most $2J$, and the force magnitude per pair is bounded by $16g/(9a^2q^2)$. For the second, all contributions to the leader remain targetward. Retaining only the block $m\ge\lceil J/2\rceil$ gives at least $2J^3$ particles, each with targetward separation at least $aJ/4$ and total separation at most $2aJ$.

Set $t_+=2\ell_0/v_J=32a^2\ell_0/(gJ)$. The choice of gap ensures $B_Jt_+\le\delta/2$. On the event that every Brownian displacement has norm at most $\ell_0/2$ up to $t_+$, the displacement stopping threshold cannot be reached: integrated drift plus Brownian displacement remains strictly below $\delta$. If no particle had been absorbed, the leader would satisfy $z_0(t_+)\le\ell_0-v_Jt_++\ell_0/2<0$, a contradiction. A union bound over the three Brownian coordinates therefore yields
\begin{equation}
\Prob(T_N>t_+)\le 6N
\exp\!\left[-\frac{g\ell_0 J}{1536Da^2}\right].
\label{eq:SM-coulomb-fast-hit}
\end{equation}
This probability is exponentially small in $J\propto N^{1/3}$, so with overwhelming probability the first hit occurs before $t_+\propto J^{-1}$.

To turn this into a bound on the mean, a heavy tail of $T_N$ must be excluded.  Since Coulomb repulsion is central and reciprocal, the results in Sec.~\ref{sec:upper-product} can be applied to achieve this. The leader and four fixed particles in the first lattice layer have initial target distances at most $Z=\ell_0+a$. The comparison in Sec.~\ref{sec:upper-product}, applied to the full system and then retaining these five factors, gives
\begin{equation}
\Prob(T_N>t)\le\operatorname{erf}\!\left(\frac{Z}{\sqrt{4Dt}}\right)^5,
\qquad \E T_N^2\le\frac{5Z^4}{\pi^2D^2}.
\label{eq:SM-coulomb-second-moment}
\end{equation}
Indeed, $\operatorname{erf}(Z/\sqrt{4Dt})\le\min\{1,Z/\sqrt{\pi Dt}\}$ and $\E T_N^2=2\int_0^\infty t\Prob(T_N>t)\,\dd t$. Five is the smallest number of factors for which the tail $\Prob(T_N>t)\lesssim t^{-5/2}$ makes this integral finite. Cauchy--Schwarz now gives the desired upper bound
$\E T_N\le t_++\sqrt{\E T_N^2\Prob(T_N>t_+)}=O(J^{-1})$.

For the lower bound, take $B_N=B_J$ and $D_N^{\rm eff}=D$ in Theorem~2. Then $t_N^\star\sim\ell_0/B_J=\Theta(J^{-1})$. On the event that all Brownian displacements up to $t_N^\star$ are at most $\delta/2$, the same stopping argument applies because $B_Jt_N^\star<\ell_0<\delta/2$. Thus the drift budget holds with failure probability at most $6N\exp[-\delta^2/(48Dt_N^\star)]\to0$, and Theorem~2 supplies the matching lower bound. Consequently,
\begin{equation}
\E T_N=\Theta(J^{-1})=\Theta(N^{-1/3}).
\label{eq:SM-coulomb-scale}
\end{equation}
This proves Example~C of the main text.  %The faster search is caused by deterministic accumulation of distant repulsive forces.

%The finite-$N$ dynamics used here is well defined from the stated initial positions. To see this directly, put $U=g\sum_{i<j}|X_i-X_j|^{-1}$. In this argument, $X$ and $W$ collect all particle positions and Brownian motions, and $\nabla_X$ and $\Delta_X$ act on the full configuration. Away from collisions, $\Delta_XU=0$ in three dimensions, and It\^o's formula gives $\dd U=-|\nabla_XU|^2\dd t+\sqrt{2D}\nabla_XU\cdot\dd W$. The nonnegative stopped process is a supermartingale, so the probability of reaching a pair distance $\epsilon$ before a fixed time is at most $U(x)\epsilon/g$. Letting $\epsilon\downarrow0$ excludes collisions for each finite $N$. On a bounded-energy event the force is bounded, also excluding finite-time spatial explosion. This establishes the noncolliding solution without any assumption about later-time density.

\section{Numerical checks}
\label{sec:numerics}

Here, we numerically compare multiple passive and constantly driven searchers with the same soft repulsion and thermal diffusivity.  The results are consistent with the ``logarithmic class'' proposed in the main text.

We simulate particles in the half-plane $x>0$, absorbed at $x=0$, with
\begin{equation}
\dd X_i=\Bigl[\sum_{j\ne i}F_{ij}-v\,e_x\Bigr]\dd t+\sqrt{2D}\,\dd W_i,\qquad
F_{ij}=\kappa\Bigl(1-\frac{r_{ij}}{R}\Bigr)_{\!+}\frac{X_i-X_j}{r_{ij}},
\label{eq:SM-numerical-model}
\end{equation}
where $r_{ij}=|X_i-X_j|$ and $e_x$ is the unit normal pointing away from the target. We take $D=0.5$, $\kappa=0.4$, and $R=1.3$, and start the particles in a row parallel to the target, $X_i(0)=(1,\,i-(N+1)/2)$. The gap and the spacing are then both one, and each particle initially has two neighbors within the interaction range. We compare passive searchers, $v=0$, with searchers driven toward the target at constant speed $v=0.8$. In both cases, Proposition~1(i), with the bounded drive added to the force budget, and the upper bound of Sec.~\ref{sec:upper-product} predict $(\ln N)\E T_N\to\ell_0^2/(4D)=1/2$.

The dynamics is integrated with Euler steps of length $\Delta t=10^{-3}$. Absorption is imposed on the Brownian-bridge interpolation of each step \cite{Gobet}: a step from $x_0>0$ to $x_1$ crosses the target with probability $\exp[-x_0x_1/(D\Delta t)]$, or with certainty if $x_1\le0$, and the crossing time is then sampled exactly \cite{MichaelSchucanyHaas}. We use $N=2^3,\ldots,2^{16}$, with $1200$ to $30000$ trajectories per value. As a consistency check of the algorithm, we checked the $\kappa=0$ (independent) case, where the scheme reproduces the exact independent means.

Figure~\ref{fig:SM-numerics} compares the simulations with independent searchers subject to the same drive, whose mean is
\begin{equation}
\E T_N^{\rm ind}=\int_0^\infty q_v(t)^N\,\dd t,\qquad
q_v(t)=\Phi\Bigl(\frac{1-vt}{\sqrt{2Dt}}\Bigr)-e^{v/D}\,\Phi\Bigl(\frac{-1-vt}{\sqrt{2Dt}}\Bigr),
\label{eq:SM-numerical-survival}
\end{equation}
where $q_v$ is the single-particle survival probability and $\Phi$ is the standard normal distribution function. The lines show $\E T_N^{\rm ind}$, evaluated up to $\ln N=250$, where $(\ln N)\E T_N^{\rm ind}$ approaches the predicted constant $1/2$. %the approach is slow because the relative corrections are of order $\ln\ln N/\ln N$. 
The symbols show the interacting simulations (evaluated up to $N=2^{16}$), which display the same $1/\ln N$ scaling. 

\begin{center}
\includegraphics[width=\linewidth]{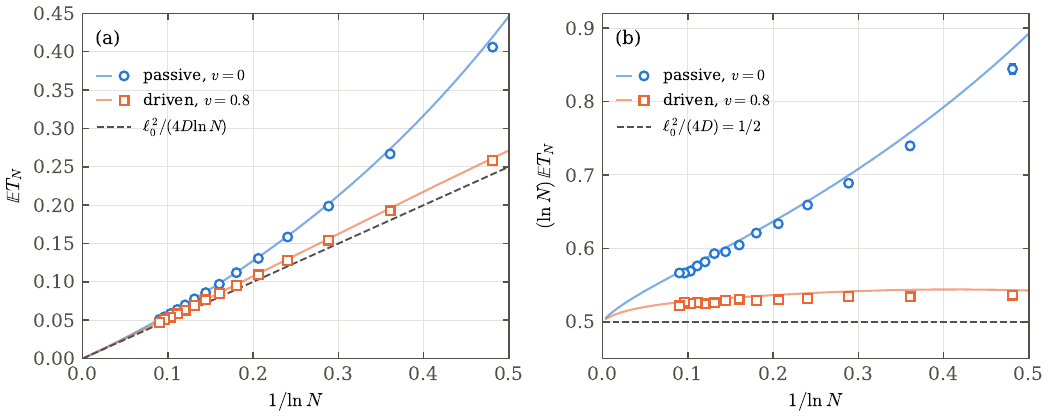}
\captionof{figure}{\label{fig:SM-numerics}Mean EFPT of the repulsive searchers of Eq.~\eqref{eq:SM-numerical-model}. Symbols: simulations for passive ($v=0$, circles) and driven ($v=0.8$, squares) searchers, with 95\% intervals. Lines: exact means of independent searchers, Eq.~\eqref{eq:SM-numerical-survival}. Dashed: the predicted asymptote. (a) $\E T_N$ against $1/\ln N$. (b) $(\ln N)\E T_N$, which tends to $\ell_0^2/(4D)=1/2$.}
\end{center}

\bibliographystyle{apsrev4-2}
\bibliography{refs}